\RequirePackage{amsmath}

\documentclass[]{interact}

\usepackage{epstopdf}
\usepackage[caption=false]{subfig}

\usepackage[numbers,sort&compress]{natbib}
\bibpunct[, ]{[}{]}{,}{n}{,}{,}
\makeatletter
\def\NAT@def@citea{\def\@citea{\NAT@separator}}
\makeatother

\usepackage{graphicx}
\usepackage{amsfonts,amssymb,enumerate}
\usepackage{array,epic,rotating,epsfig} 
\usepackage{graphics}
\usepackage{todonotes}
\usepackage{mathtools}

\usepackage{tikz,pgfplots}
\pgfplotsset{compat=1.16}
\usetikzlibrary{decorations.text}
\usetikzlibrary{plotmarks}
\usetikzlibrary{shapes}
\usetikzlibrary{trees}
\usetikzlibrary{calc}
\usetikzlibrary{patterns}
\usetikzlibrary{math}
\usepgfplotslibrary{fillbetween}
\usepgfplotslibrary{colorbrewer}

\usetikzlibrary{arrows,intersections,calc,patterns,shapes,arrows,positioning,fillbetween}
\makeatletter 
\pgfdeclarepatternformonly[\LineSpace,\tikz@pattern@color]{my north east lines}{\pgfqpoint{-1pt}{-1pt}}{\pgfqpoint{\LineSpace}{\LineSpace}}{\pgfqpoint{\LineSpace}{\LineSpace}}%
{
	\pgfsetcolor{\tikz@pattern@color} 
	\pgfsetlinewidth{0.4pt}
	\pgfpathmoveto{\pgfqpoint{0pt}{0pt}}
	\pgfpathlineto{\pgfqpoint{\LineSpace + 0.1pt}{\LineSpace + 0.1pt}}
	\pgfusepath{stroke}
}
\makeatother 
\newdimen\LineSpace
\tikzset{
	line space/.code={\LineSpace=#1},
	line space=3pt
}
\usepgfplotslibrary{fillbetween}
\usetikzlibrary{graphs} 
\usetikzlibrary{arrows}
\usetikzlibrary{matrix}
\usetikzlibrary{quotes}

\usepackage{pgf}
\usetikzlibrary{spy}
\usetikzlibrary{backgrounds}
\usetikzlibrary{decorations}

\usepackage[section]{algorithm}
\usepackage{algorithmic} 
\usepackage{longtable,multirow} 
\usepackage{url} 
\usepackage{subfig}
\usepackage{cancel}

\newcommand {\setJ}{J}

\newcommand {\N}{\mathbb{N}}

\newcommand {\R}{\mathbb{R}}

\renewcommand{\b}{\mathcal S}
\newcommand {\E}{\mathcal{E}}

\newcommand {\I}{\mathcal{I}}

\newcommand{\M}{\mathcal{M}}

\newcommand {\U}{\mathcal{U}}

\newcommand {\X}{\mathcal{X}}
\newcommand {\Y}{\mathcal{Y}}
\newcommand {\bigo}{\mathcal{O}}
\newcommand {\NP}{\mathcal{NP}}

\newcommand {\Xeff}{\mathcal{X}_{\makebox{\scriptsize{\it E}}}}
\newcommand {\Ynd}{\mathcal{Y}_{\makebox{\scriptsize{\it N}}}}
\newcommand {\Ywnd}{\mathcal{Y}_{\makebox{\scriptsize{\it wN}}}}
\newcommand {\Xweff}{\mathcal{X}_{\makebox{\scriptsize{\it wE}}}}

\DeclareMathOperator {\conv}{conv}	

\newcommand {\eps}{\varepsilon}

\newcommand{\esa}{ESA} 
\newcommand{\ce}{CE} 
\newcommand{\DP}{DP} 
\newcommand{\nc}{nc} 

\newcommand {\momp}{MOMP}			

\newcommand {\bbmp}{BBMP}			

\newcommand{\algorithmicinput}{\textbf{Input:}}
\newcommand{\INPUT}{\item[\algorithmicinput]}
\newcommand{\algorithmicoutput}{\textbf{Output:}}
\newcommand{\OUTPUT}{\item[\algorithmicoutput]}

\newcommand{\hide}[1]{}

\theoremstyle{plain}
\newtheorem{theorem}{Theorem}[section]
\newtheorem{lemma}[theorem]{Lemma}
\newtheorem{corollary}[theorem]{Corollary}

\theoremstyle{definition}
\newtheorem{definition}[theorem]{Definition}
\newtheorem{example}[theorem]{Example}

\theoremstyle{remark}

\begin{document}


\title{Biobjective Optimization Problems on Matroids with Binary Costs}

\author{
\name{Jochen~Gorski\textsuperscript{a} and Kathrin~Klamroth\textsuperscript{b} and Julia~Sudhoff\textsuperscript{c}\thanks{CONTACT Julia~Sudhoff. Email: sudhoff@math.uni-wuppertal.de}}
\affil{\textsuperscript{a}TH Nürnberg, Nürnberg, Germany; \textsuperscript{b,c}University of Wuppertal, Wuppertal, Germany}
}

\maketitle

\begin{abstract}
Like most multiobjective combinatorial optimization problems, biobjective
optimization problems on matroids are in general intractable and their corresponding decision problems are in general NP-hard.
In this paper, we consider biobjective optimization problems on matroids where one of the objective functions is restricted to binary cost coefficients. We show that in this case the problem has a connected efficient set with respect to
a natural definition of a neighborhood structure and hence, can be solved efficiently using a neighborhood search approach. 
This is, to the best of our knowledge, the first non-trivial problem on matroids where connectedness of the efficient set can be established.

The theoretical results are validated by  numerical experiments with biobjective minimum spanning tree problems (graphic matroids) and with biobjective knapsack problems with a cardinality constraint (uniform matroids). 
In the context of the minimum spanning tree problem, coloring all edges with cost $0$ green and all edges with cost $1$ red leads to an equivalent problem where we want to simultaneously minimize one general objective and the number of red edges (which defines the second objective) in a Pareto sense.
\end{abstract}


\section{Introduction}
Optimization problems on matroids have been frequently studied in the literature. Early references date back to the middle 1930's, see, for example, \cite{whit:onth:35}. A well-known example are \emph{graphic matroids}, i.e., minimum spanning tree problems in simple connected graphs. Single objective optimization problems on matroids can be solved efficiently by a simple greedy strategy. We refer to the books of Kung \cite{kung:asou:86} and Oxley \cite{oxle:matr:92} for a more detailed introduction into this field.

While the literature on single objective matroid optimization is relatively rich, the work on multiobjective optimization on matroids mostly focuses on multiobjective spanning tree problems. See, for example, Ruzika and Hamacher \cite{ruzi:asur:09} for a survey and Benabbou and Perny \cite{bena:onpo:15} for a more recent reference on this topic. Evolutionary methods for multiobjective spanning tree problems were suggested, among others, in Zhou and Gen \cite{zhou:gene:1999}, Knowles and Corne \cite{know:acom:2001},  Neumann and Witt \cite{neum:mult:2010} and Bossek et al. \cite{boss:onth:2019} as well as references therein. Loera at al.\ \cite{loer:comp:2010} describe heuristic approaches to general multiobjective matroid optimization problems that rely on adjacency relations and nonlinear scalarizations. The methods are implemented in the MOCHA software package \cite{leor:mocha:2009}. Approximation schemes were suggested, for example, in Grandoni et al.\ \cite{gran:newa:2014} and Bazgan et al. \cite{bazg:thep:2019}.

Multiobjective optimization problems on matroids are a special case of \emph{multiobjective combinatorial optimization} (MOCO) problems which are known to be notoriously hard. We refer to \cite{figu:easy:2017} for a recent discussion of the prevalent difficulties in MOCO problems. The decision problem of multiobjective matroid optimization is proven to be $\NP$-complete in general, see  \cite{ehrg:onma:96}.
For multiobjective spanning tree problems it was shown in \cite{hama:onsp:94} that already in the biobjective case the cardinality of the non-dominated set may grow exponentially with the size of the instance. This result applies also to multiobjective optimization problems on matroids. As a consequence, for such instances the complete enumeration of the non-dominated set is impractical since it requires an exponential amount of time. B{\"o}kler et al.\  \cite{boek:outp:2017} recently suggested to consider the concept of output sensitive complexity in the context of MOCO problems and analysed various problem classes. 
In the dissertation of B{\"o}kler \cite{boek:diss:2018} the output sensitive complexity of the biobjective spanning tree problem was related to that of biobjective unconstrained combinatorial optimization (BUCO) which is, however, also still open. 
Despite the general intractability of multiobjective spanning tree problems, it was shown in the dissertation of Seipp \cite{seipp:diss:2013} that the number of extreme supported non-dominated outcome vectors grows only polynomially with the size of the instance.

The above mentioned hardness results usually refer to MOCO instances with 'large' cost coefficients that may grow exponentially with the instance size. For problems with 'small' cost coefficients the situation is different. When coefficients are small, then the ranges of possible outcome values are bounded, which limits the size of the non-dominated set. For example, the biobjective minimum spanning tree problem has only supported efficient solutions when all  cost coefficients take only values from the set $\{0,1,2\}$, see again  \cite{seipp:diss:2013}. This implies that all efficient solutions of this problem are connected, i.e., the complete efficient set can be generated by only performing simple swap operations (e.g., pivot operations in an associated linear programming formulation) among efficient solutions. In the same work, \cite{seipp:diss:2013} show that tri-objective optimization problems on uniform matroids with one general cost function and two binary cost functions have a connected efficient set.
However, in general even comparably simple problems like BUCO may possess a non-connected efficient set, see  \cite{gors:conn:06}.

In this paper we focus on 
biobjective optimization problems on matroids that have binary  coefficients in \emph{one} of the objectives. 
While the first objective may take arbitrary non-negative integer values, we assume that the second objective takes only values from the set $\{0,1\}$. Note that binary coefficients allow for an alternative interpretation of the problem: When associating a cost of $0$, for example, with the color 'green', and a cost of $1$ with the color 'red', then we are interested in the simultaneous minimization of the cost of a solution  (w.r.t.\ the first objective) \emph{and} of the number of its red elements. 

A related problem is the \emph{multicolor matroid problem} that was discussed by Rendl and Leclerc \cite{rend:amul:88} and by Brezovec et al.\ \cite{brez:amat:88}.  
In this problem, a minimum cost solution is sought that does not exceed a given bound on the number of elements from different colors. 
Srinivas \cite{srin:matr:95} extended the results from Brezovec et al.\ \cite{brez:amat:88} to the case that the number of elements of different colors is constrained by linear inequalities. 
Hamacher and Rendl \cite{hama:colo:91} generalized the multicolor matroid  problem to combinatorial optimization problems, now allowing for elements having more than one color. Similar to \cite{brez:amat:88} the goal is to find minimum cost solutions not exceeding given bounds on the number of elements in each color. 
A different optimization objective was considered in Climaco et al.\  \cite{clim:onth:10}, who discussed a biobjective minimum cost / minimum label spanning tree problem in a graph where each edge is associated with a cost value and a label (i.e., a color). While the first objective is a classical cost objective that is to be minimized, the second objective is to find a solution with a minimal number of \emph{different} labels (i.e., colors). Since it is already $\NP$-hard to determine the minimum label spanning tree on a given graph due to a result of Chang and Leu \cite{chazn:them:97}, this problem is also $\NP$-hard.

From an application point of view, MOCO problems with one general objective function and one (or several) binary objectives are closely related to $k-\max$ optimization where the $k$th largest cost coefficient of a solution vector is to be minimized. Such problems can be translated into a series of problems with binary sum objectives in a thresholding framework, see, e.g, \cite{gors:onkm:09} for more details.

\paragraph*{Contribution.}  
This paper extends results from Chapter~10 of the dissertation of Gorski \cite{gorski:diss:2010}. It is shown that the non-dominated set of biobjective optimization problems on matroids with one general and binary objective function contains only supported efficient solutions and is connected. This is the foundation for an efficient exact algorithm that enumerates the non-dominated set using a neighborhood search approach, i.e., using simple swaps between elements contained in different (efficient) bases of the problem. This \emph{Efficient Swap Algorithm} \esa\ can be interpreted as an extension of the algorithm of \cite{gabo:effi:84} for a constrained version of the problem that is guaranteed to generate the complete non-dominated set.
To the best of our knowledge, this is the first non-trivial optimization problem on matroids for which connectedness of the efficient set is  established. 

\paragraph*{Organization of the paper.}
The remainder of this paper is organized as follows. In Section~\ref{sec:prelimatro} we recall basic concepts from matroid theory and from multiobjective optimization that are relevant for the subsequent sections. The biobjective matroid optimization problem with one binary cost objective is introduced in Section~\ref{sec:bbmp}. The neighborhood search algorithm \esa\  is presented in Section~\ref{sec:solvebbmp}, and connectedness of the efficient set is proven in Section~\ref{s:basescon}.  The numerical results presented in Section~\ref{sec:bbmpnumres} confirm the efficiency of the algorithm introduced in Section~\ref{sec:solvebbmp}.
The paper is concluded in Section~\ref{sec:bbmpcon} with some ideas for future research.

\section{Matroid and Multiobjective Optimization Preliminaries}\label{sec:prelimatro}

We first review basic concepts from matroid theory and multiobjective optimization. 
For more details on matroid theory we refer to the books of Kung \cite{kung:asou:86} and Oxley \cite{oxle:matr:92}. For an introduction into the field of multiobjective optimization, see, e.g., the books of Ehrgott \cite{ehrg:mult:05} and Miettinen \cite{miet:nonl:99}.

\subsection{Matroids}\label{subsec:matroids}
Let $\E=\{e_1,\ldots,e_n\}$ be a finite \emph{ground set} with $n\in\N$ elements and let $\I$ be a subset of the power set $\mathcal{P}(\E)$ of $\E$. The ordered pair $\M=(\E,\I)$ is called a \emph{matroid} if the following three conditions are satisfied: 
\begin{align}
& \emptyset\in\I \tag{M1}\label{eq:M1}\\
& (I\in \I \;\wedge\; I'\subseteq I) \quad\Rightarrow\quad I'\in \I \tag{M2}\label{eq:M2}\\
& \forall I_1,I_2\!\in\I \text{~with~} |I_1|\!<\!|I_2|\;\; \exists \,e\!\in\! I_2\setminus I_1\;\colon\;
    I_1\cup\{e\}\in \I . \tag{M3}\label{eq:M3}
\end{align} 
$|I|$ denotes the cardinality of a finite set $I$. If $\M=(\E,\I)$ is a matroid, then all sets $I\in \I$ are called \emph{independent sets}. Conversely, a  subset of $\E$ is called \emph{dependent} if it is not contained in $\I$. 

An independent set $I\in\I$ is called \emph{maximal} when $I\cup \{e\}\not\in \I$ for all $e\in \E\setminus I$. 
Similarly, a dependent set  $D\in\mathcal{P}(\E)\setminus\I$ is called \emph{minimal} if  $D\setminus\{e\}\in\I$ for all $e\in D$. 
Maximal independent sets are called \emph{bases} of the matroid,  and minimal dependent sets are called \emph{circuits} of the matroid.
All bases of a matroid have the same cardinality, which is referred to as the \emph{rank} of $\M$. We denote the set of all bases of a given matroid by $\X$. 

Given a matroid $\M=(\E,\I)$, a basis $B\in\X$, and an element $e\in\E\setminus B$, then $B\cup\{e\}$ contains a uniquely determined circuit $C(e,B)$ 
containing $e$. This circuit is also called the \emph{fundamental circuit} of $e$ w.r.t.\ $B$.
An important property of matroids is the \emph{basis exchange property}:
\begin{align}\tag{B}\label{eq:B}
    & \forall E,F\!\in\X\;\; \forall e\!\in\! E\setminus F\;\; \exists \,f\!\in\! F\setminus E\;\colon\;
    (E\cup\{f\})\setminus\{e\} \in \X .
\end{align}
The following stronger version of the basis exchange property was proven in \cite{brua:comm:69}.
\begin{lemma}[\cite{brua:comm:69}]\label{l:baseex}
Let $E,F\in\X$. For all $e\in E\setminus F$ there exists $f\in F\setminus E$ such that both $(E\cup\{f\})\setminus\{e\}$ and $(F\setminus\{f\})\cup\{e\}$ are bases in $\X$.
\end{lemma}
In this context, two bases of a matroid are called \emph{adjacent} if they have $m-1$ elements in common, assuming that the matroid is of rank $m$. According to the basis exchange property \eqref{eq:B} and Lemma~\ref{l:baseex}, a given basis can be transformed into an adjacent basis by exactly one basis exchange. We refer to this as a \emph{swap operation} in the following.

If a subset $S\subseteq \E$ of the ground set $\E$ is deleted from $\E$, we obtain the \emph{restriction} of $\M$ to $\E\setminus S$. The ground set of this matroid is the set $\E\setminus S$ and its independent sets are those independent sets of $\M$ that are completely contained in $\E\setminus S$, i.e., that do not contain any elements from $S$. We write $\M-S$ for short.

Moreover, if an independent set $I$ of $\M$ is \emph{contracted} we obtain the \emph{contraction} of $\M$ to $I$ denoted by $\M/I$. The ground set of $\M/I$ is given by $\E\setminus I$, and its independent sets are the sets $I'\subseteq (\E\setminus I)$ such that $I'\cup I$ is an independent set of $\M$.

A classical example for a matroid is the \emph{uniform matroid} of rank $k$, denoted by $\U_{k,n}$. The independent sets of $\U_{k,n}$ are all subsets of $\E=\{e_1,\ldots,e_n\}$ that have at most $k$ elements, and the bases of $\U_{k,n}$ are all subsets of $\E$ that have exactly $k$ elements. A subset of $\E$ is a circuit of $\U_{k,n}$ if it contains exactly $k+1$ elements of $\E$. 
Another common example is the  \emph{graphic matroid}.  Given a finite undirected graph $G=(V,E)$ with node set $V$ and edge set $E$, the independent sets of $\M(G)$ are all forests in $G$, and the bases of $\M(G)$ are all spanning forests of $G$. When $G$ is connected, then $\X$ is the set of all spanning trees of $G$. In this case, a circuit is referred to as a cycle. It is easy to verify that uniform matroids and graphic matroids satisfy the conditions \eqref{eq:M1}, \eqref{eq:M2} and \eqref{eq:M3}. We will use graphic matroids to illustrate the results throughout this paper.

Since the efficiency of the methods developed in this paper depends on the structure of the considered matroid, we briefly review some further matroids in the following. First consider the \emph{matching matroid} that is also
defined on a finite undirected graph $G=(V,E)$. The ground set of the matching matroid is a subset of the vertices of $G$, i.e.,  $\E\subseteq V$, and all subsets of $\E$ that can be covered by a matching of $G$ are independent.
A special case of the matching matroid is the \emph{transversal matroid}, that is a matching matroid on a bipartite graph $G=(V_1\cup V_2, E)$ with bipartition $V=V_1\cup V_2$, where $\E$ equals $V_1$ or $V_2$. The \emph{partition matroid} is defined on a groundset $\E=\{e_1,\dots,e_n\}$ of $n$ elements that is partitioned into $p\geq 1$ subsets $E_i$, $i=1,\dots,p$. For given non-negative bounds $b_i\geq 0$, $i=1,\dots,p$, a set $I\subseteq\E$ is independent whenever $\vert E_i\cap I\vert\leq b_i$ for all $i=1,\dots,p$. Note that the uniform matroid is a special case of the partition matroid with $p=1$.

\subsection{Multiobjective Optimization}\label{subsec:MOP}

Now suppose that a matroid $\M=(\E,\I)$ is given and that $p\geq2$ cost coefficients $w_i(e)\geq 0$, $i=1,\ldots,p$, are associated with each element $e\in\E$ of the ground set $\E$. The cost of a subset $S\subseteq\E$ in the $i$th objective is computed as $w_i(S)=\sum_{e\in S} w_i(e)$, $i=1,\ldots,p$.   
Then the \emph{multiple objective matroid problem} ({\momp}) can be formulated as
\begin{align*}\label{p:momp}\tag{MOMP}
\min\; & w(B)\,=\, (w_1(B), \ldots , w_{p}(B))\\
\text{s.t. } & B \in \X.
\end{align*}
The feasible solutions of~\eqref{p:momp} are the bases $B\in\X$ of the matroid, and $y=w(B)\in\R^p$ denotes the \emph{cost vector} or \emph{outcome vector}  of the basis $B$. In the following, we will enumerate different outcome vectors by using superscripts and refer to their components by subscripts. The minimization in problem \eqref{p:momp} is understood w.r.t.\ the Pareto concept of optimality that is based on the componentwise ordering in $\R^p$:
\begin{align*}
    y^1\leqq y^2 &:\Leftrightarrow y_i^1 \leq y_i^2,\text{ }i=1,\ldots ,p,\\
    y^1\leqslant y^2 &:\Leftrightarrow y_i^1\leq y_i^2,\text{ }i=1,\ldots,p\text{ and }y^1\neq y^2,\\
    y^1<y^2 &:\Leftrightarrow y_i^1<y_i^2, \text{ }i=1,\ldots,p.
\end{align*}
We say that an outcome vector $y^1$ \emph{dominates} another outcome vector $y^2$ if and only if $y^1\leqslant y^2$, and $y^1$ \emph{strongly dominates} $y^2$ if and only if $y^1<y^2$. 
A feasible solution $B\in\X$ (i.e., a feasible basis) is called \emph{efficient} or \emph{Pareto optimal} if there does not exist another feasible solution $\bar{B}\in\X$ that \emph{dominates} $B$, i.e., for which $w(\bar{B})\leqslant w(B)$. Similarly, $B\in\X$ is called \emph{weakly efficient} or \emph{weakly Pareto optimal} if there does not exist $\bar{B}\in\X$ that \emph{strongly dominates} $B$, i.e., for which $w(\bar{B})< w(B)$ (cf.\ also Figure~\ref{f:zfktraum} in Section~\ref{s:basescon} below). We are interested in finding the \emph{efficient set} (or the \emph{weakly efficient set}, respectively) of problem \eqref{p:momp} given by 
\begin{align*}
    \Xeff\coloneqq \{B\in\X:\text{there exists no }\bar{B}\in\X\text{ with }w(\bar{B})\leqslant w(B)\},\\
    \Xweff\coloneqq \{B\in\X:\text{there exists no }\bar{B}\in\X\text{ with }w(\bar{B})< w(B)\}.
\end{align*}
Given $\X_E$ and $\X_{wE}$, the images of these two sets under the vector-valued mapping $w$ are called \emph{non-dominated set} and \emph{weakly non-dominated set}, respectively:
\begin{align*}
    \Ynd\coloneqq w(\Xeff)\\
    \Ywnd\coloneqq w(\Xweff).
\end{align*}
A subset $\X_{cE}$ of $\Xeff$ satisfying $w(\X_{cE})=\Ynd$ is called a \emph{complete set of efficient solutions}. Note that in general $\X_{cE}\subsetneq\Xeff$. If in addition $|w(\X_{cE})|=|\Ynd|$ holds true, we say that the set $\X_{cE}$ is of minimal cardinality or just minimal, for short. Note that in this case, $\X_{cE}$ contains exactly one efficient solution for each vector in the non-dominated set. Algorithms designed to solve~\eqref{p:momp} often aim to compute $\Ynd$ and $\X_{cE}$ rather than $\Ynd$ and $\Xeff$.
An efficient basis $B$ is called \emph{supported efficient} if it is a minimizer of the non-trivial weighted sum problem $\min\{\sum_{i=1}^p\lambda_i w_i(B),\, B\in\X\}$ with $\lambda_i\in (0,1)$, $i=1,\ldots,p$ and  $\sum_{i=1}^p\lambda_i=1$.
Note that the image $w(B)$ of a supported efficient basis $B$ is called supported non-dominated outcome vector and lies on the boundary of the convex hull $\conv(\Y)$ of the set $\Y=w(\X)$ of feasible outcome vectors in the objective space.
Moreover, if a basis $B$ is supported efficient and if $w(B)$ is an extreme point of $\conv(\Y)$ then $B$ is called an \emph{extreme supported efficient basis} and $w(B)$ is called an \emph{extreme supported non-dominated point}. See Figure~\ref{f:zfktraum} in Section~\ref{s:basescon} for an illustration.

An important subset of the efficient set is the set of lexicographically optimal solutions:
An outcome vector $y^1$ is \emph{lexicographically optimal} if for all other outcome vectors $y^2$ it holds that $y^1_i<y^2_i$ with $i=\min\{j\in\{1,\dots,p\}:y^1_j\neq y^2_j\}$.

Based on the concept of adjacent bases, the \emph{adjacency graph} $G=(V,E)$ of efficient bases of Problem~\eqref{p:momp} is defined analogous to \cite{gors:conn:06}. The node set $V$ consists of all efficient bases of~\eqref{p:momp}. An (undirected) edge is introduced between all pairs of vertices corresponding to adjacent bases of the underlying problem. These edges form the set $E$. The set $\Xeff$ is
said to be connected if its corresponding adjacency graph $G$ is connected, i.e., if every pair of vertices in $V$ is connected by a path. As shown in~\cite{gorski:diss:2010}, the adjacency graph $G$ is not connected in general, even if it is extended to include weakly efficient bases. Nevertheless, the adjacency graph always contains a connected component given by the supported efficient bases of~\eqref{p:momp}, see \cite{ehrg:onma:96}. Although the adjacency graph is not connected in general, many solution methods make use of the adjacency structure of matroids. Some examples for such solution strategies are described in \cite{loer:comp:2010}.

\section{Problem Formulation and Notation}\label{sec:bbmp}

Let $\M=(\E,\I)$ be a matroid and let $\X$ denote  the set of all bases of $\M$. We assume that $\textnormal{rank}(\M)=m>0$, i.e., the cardinality $|B|$ of all bases  $B\in\X$ is equal to $m$. In the following we consider two different types of cost functions on the ground set $\E$. While the first function $c:\E\to\N$ is given by arbitrary non-negative integer coefficients, we assume that the second cost function $b:\E\to\{0,1\}$ only takes binary values on the elements of the ground set. According to these definitions the two different costs of a basis $B\in\X$ are given by $c(B)=\sum_{e\in B}c(e)$ and $b(B)=\sum_{e\in B}b(e)$, respectively. The related \emph{biobjective matroid problem with binary costs} ({\bbmp}) is given by

\begin{equation}\label{p:bbmp}\tag{$BBMP$}
\min_{B\in \X}\left(c(B),b(B)\right).
\end{equation}

Since the second cost function $b$ has binary coefficients for all elements $e\in\E$, the corresponding objective function values of feasible bases $B\in\X$ are lower bounded by zero and upper bounded by $m$. In other words, $b(\X)\coloneqq \{b(B):\,B\in \X\}\subseteq\{0,\ldots,m\}$ is of size $\bigo(m)$, and thus the same bound also holds for $\Ynd$.

For solving Problem~\eqref{p:bbmp} we introduce the following two associated $\eps$-constraint versions of the problem. The first is given by

\begin{equation}\label{p:bbmpleq}\tag{$BMP_\leq$}
\arraycolsep2pt
\begin{array}{lrrc}
&\multicolumn{3}{l}{\min c(B)}\\
\mbox{s.t.}&b(B)& \leq&k,\\
	&B&\in &\X,
\end{array}
\end{equation}
where $k\in\{0,\ldots,m\}$ is a fixed integer bound on the binary cost function $b$. From the theory of multiple criteria optimization (see e.g.~\cite{chan:mult:83}) we know that each optimal solution of Problem~\eqref{p:bbmpleq} is at least weakly efficient for Problem~\eqref{p:bbmp}. Note that this is not true in general when the inequality constraint in Problem~\eqref{p:bbmpleq} is replaced by an equality constraint. Indeed, given an optimal solution of the equality constrained problem
\begin{equation}\label{p:bbmpeq}\tag{$BMP_=$}
\arraycolsep2pt
\begin{array}{lrrc}
&\multicolumn{3}{l}{\min c(B)}\\
\mbox{s.t.}&b(B)& =&k,\\
	&B&\in &\X,
\end{array}
\end{equation}
this solution may be dominated in Problem~\eqref{p:bbmp}. An example for this situation can be seen in Figure~\ref{f:zfktraum}. There, each basis $B\in\X$ that maps to the outcome vector $(c(B),b(B))=(18,5)$ is optimal for Problem~\eqref{p:bbmpeq} with $k=5$, while it is dominated by all bases that map to the outcome vector $(17,4)$ for the biobjective problem. Nevertheless, we will use Problem~\eqref{p:bbmpeq} to generate a sequence of optimal solutions by varying $k\in\{0,\ldots,m\}$ and show that there exists a critical index $j$ such that for all $k\leq j$ all generated bases that are optimal for Problem~\eqref{p:bbmpeq} correspond to efficient bases of Problem~\eqref{p:bbmp}. 

Note that the binary cost function $b$ introduced above also allows for another interpretation as used, for example, in \cite{gabo:effi:84} and~\cite{gusf:matr:84}: Given a matroid $\M=(\E,\I)$ and a (first) cost function $c:\E\to\N$, one of the two colors red and green is assigned to each element of $\E$. In \cite{gabo:effi:84} and~\cite{gusf:matr:84} algorithms are presented that determine a minimum cost basis $B\in\X$ that contains exactly $k$ red elements from $\E$ (here, $k$ is a predetermined parameter). To establish a connection between the problem discussed in~\cite{gabo:effi:84} and~\cite{gusf:matr:84} and the problems considered here, we simply identify the red elements $r\in\E$ from the ground set $\E$ with the binary costs $b(r)=1$, while all green elements $g\in\E$ are considered to have binary cost $b(g)=0$. Hence, determining a minimum cost basis $B\in\X$ containing at most or exactly $k$ red elements from $\E$ corresponds to solving Problem~\eqref{p:bbmpleq} and~\eqref{p:bbmpeq}, respectively. In this context, especially Problem~\eqref{p:bbmpeq} can be seen as a generalized version of a single objective matroid problem with an additional constraint, where the original problem is obtained when $\E$ only consists of red elements and $k=m$. Note that for a better illustration, we will make use of the idea of red and green elements in the further sections.

\section{Solving Biobjective Matroid Problems with Binary Costs}\label{sec:solvebbmp}

In this section we present an algorithm that computes the complete non-dominated set of Problem~\eqref{p:bbmp} in polynomial time. The method is based on the ideas stated in~\cite{gabo:effi:84} and can be used to establish a connectedness result for the adjacency graph of Problem~\eqref{p:bbmp}. In more detail, the algorithm generates a sequence of optimal solutions of Problem~\eqref{p:bbmpeq} for decreasing right-hand side values $k$. In Subsection~\ref{subsec:baseswaps} we formulate the theoretical results that are needed to prove the correctness of the method, and in Subsection~\ref{sec:bbmpgabo} we present the algorithm itself and illustrate it at a graphic matroid.

To simplify the discussion, we use the following notation for set operations  throughout this and the following sections: Let $S$ denote a subset of the finite ground set $\E$ and let $e,f\in\E$. We write $S+e$ to denote the set $S\cup\{e\}$ and  $S-f$ to denote the set $S\setminus\{f\}$. Furthermore, let $S^c\coloneqq \E\setminus S$ denote the complement of $S$ in $\E$. To further simplify the notation we assume throughout this section that set operations are executed from left to right. 
Given an instance of Problem~\eqref{p:bbmp}, we denote by $E_0\coloneqq \{e\in\E: b(e)=0\}$ the subset of $\E$ containing all elements with binary cost $0$ (green elements) while $E_1\coloneqq \{e\in\E: b(e)=1\}=E_0^c$ denotes the set of elements with binary cost $1$ (red elements). By definition, $E_0$ and $E_1$ form a partition of $\E$.

Throughout this section, we consider Problem \eqref{p:bbmp} on a given matroid $\M=(\E,\I)$ with set of feasible bases $\X$.

\subsection{Minimal Swaps}\label{subsec:baseswaps}
The idea of our approach to generate the complete non-dominated set of Problem~\eqref{p:bbmp} is based on the stronger version of the basis exchange property for matroids stated in Lemma~\ref{l:baseex}. Given this property we define \emph{swaps} between elements from $E_0$ and $E_1$. 

\begin{definition}\label{d:matswap}
Let $B\in\X$. Then the \emph{swap} $(e,f)$   w.r.t.\ $B$ is an ordered pair of elements such that $e\in E_1\cap B$, $f\in E_0\setminus B$ and $B-e+f\in\X$ is a basis. The \emph{cost} of the swap $(e,f)$ is defined as  $c(e,f)\coloneqq c(f)-c(e)$. A swap $(e,f)$ is called \emph{minimal} w.r.t.\ $B$ if $c(e,f)\leq c(e^\prime,f^\prime)$ for all $e^\prime\in E_1\cap B$ and $f^\prime\in E_0\setminus B$ with $B-e^\prime+f^\prime\in\X$. 
\end{definition}

By definition, a swap always improves the binary cost function by one unit since a red element from $E_1$ is replaced by a green element from $E_0$. The idea of the \emph{efficient swap algorithm} (ESA) is to generate a sequence of minimal swaps that yields all non-dominated outcome vectors of Problem~\eqref{p:bbmp}, as outlined in Algorithm~\ref{a:solvebbmpGeneral}.

\begin{algorithm}
\caption{Outline of the Efficient Swap Algorithm (\esa) for Biobjective Matroid Problems with one Binary Cost Function}\label{a:solvebbmpGeneral}
\begin{algorithmic}[1]
\INPUT An instance $((\M,\X,(c,b))$ of Problem~\eqref{p:bbmp}.
\OUTPUT $\Ynd$ and a complete set $\X_{cE}$ of efficient solutions. 
\STATE Determine a basis $B_{j}$, which is optimal with respect to $c$, and a basis $B_u$, which is optimal with respect to $b$, such that both bases have as many elements as possible in common.
\STATE Compute a sequence of minimal swaps, which describes the necessary swaps to get from basis $B_j$ to basis $B_u$.
\STATE Sort the swaps in non-decreasing order with respect to their costs.
\STATE Compute from the sorted swap sequence a complete set $\X_{cE}$ of efficient solutions and the corresponding outcome vectors $\Ynd$.
\RETURN $\X_{cE}$ and $\Ynd$.
\end{algorithmic}
\end{algorithm}

A detailed description of this approach will be given in Algorithm~\ref{a:solvebbmp} below, after a thorough analysis of the individual steps.

For this purpose, let $i\in\{0,\ldots,m\}$ and $\X_i\coloneqq \{B\in\X:\,|B\cap E_0|=i\}$ be the set of all bases with exactly $i$ green elements. Note that $\X_i$ might be empty for low or high values of $i$, respectively. Furthermore, let
\[\b_i\,\coloneqq \,\{B\in\X_i:\,c(B)\leq c(B^\prime)\,\forall B^\prime\in\X_i\}\]
denote the set of all bases with minimal costs containing exactly $i$ green elements from $E_0$. By construction, $B\in\b_i$ is an optimal basis of Problem~\eqref{p:bbmpeq} with right hand side value $k=m-i$.

From \cite{gabo:effi:84} we recall that given an optimal solution $B\in\b_{i-1}$, a minimal swap can be used to generate an optimal solution contained in $\b_i$ whenever $\b_i$ is non-empty.

\begin{theorem}[see \cite{gabo:effi:84}, Augmentation Theorem 3.1]\label{t:minbaseswap}
Let $B\in\b_{i-1}$ for an $i\in\{1,\ldots,m\}$ and assume that $\b_{i}\neq\emptyset$. If the swap $(e,f)$ is minimal w.r.t.\ $B$, then $B-e+f$ is contained in $\b_i$. 
\end{theorem}
The following result is an immediate consequence of Theorem~\ref{t:minbaseswap}.

\begin{corollary}\label{c:minmaxswap}
Let $s,t\in\N$ with $0\leq s<t\leq m$ such that $\b_s\neq\emptyset\neq\b_t$. Then $\b_i\neq\emptyset$ for all $i\in\{s,\ldots,t\}$.
\end{corollary}

\begin{figure}
\begin{center}

\begin{tikzpicture}
\node[draw,circle,thick,fill=black!10] (1) at (0,0)[] {1};
\node[draw,circle,thick,fill=black!10] (2) at (2,0) {2};
\node[draw,circle,thick,fill=black!10] (3) at (4,0) {3};
\node[draw,circle,thick,fill=black!10] (4) at (0,-2) {4};
\node[draw,circle,thick,fill=black!10] (5) at (2,-2) {5};
\node[draw,circle,thick,fill=black!10] (6) at (4,-2) {6};
\node[draw,circle,thick,fill=black!10] (7) at (6,-1) {7};


\graph {
(1) --["1",thick,draw=black!40!red] (2);
(2) --["2",thick,draw=black!40!red] (3);
(1) --["4",thick,swap,draw=black!40!red] (4);
(2) --["2",thick,swap,draw=black!40!red] (4);
(4) --["8",dashed,draw=black!50!green,thick,swap] (5);
(2) --["9",dashed,thick,swap,draw=black!50!green] (5);
(5) --["3",thick,swap,draw=black!40!red] (6);
(2) --["7",thick,dashed,draw=black!50!green] (6);
(3) --["5",thick,dashed,draw=black!50!green] (6);
(3) --["4",thick,dashed,draw=black!50!green] (7);
(6) --["6",thick,swap,draw=black!40!red] (7);
};
\end{tikzpicture}

\end{center}
\caption{Graph $G=(V,E)$ with costs $c(e)$, $e\in E$, for the graphic matroid considered in Example~\ref{ex:bbmpswap}. Dashed green lines correspond to edges $e\in E$ with $b(e)=0$ and solid red lines correspond to edges with $b(e)=1$, respectively.}\label{f:bbmpex}
\end{figure}
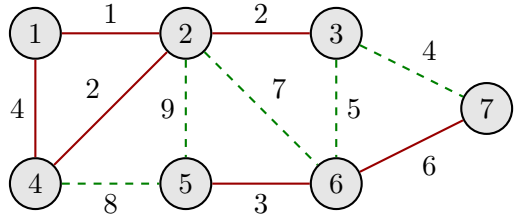

Note that Corollary~\ref{c:minmaxswap} does not state that Problem~\eqref{p:bbmpeq} is feasible for all right-hand side values $k\in\{0,\ldots,m\}$. However, it implies that there exist fixed lower and upper bounds $l,u\in\N$ (satisfying $0\leq l\leq  u\leq m$) such that $\b_i\neq\emptyset$ for all $i\in\{l,\ldots,u\}$ while $\b_j=\emptyset$ for all $j\in\{0,\ldots,m\}\setminus\{l,\ldots,u\}.$

The results of Theorem~\ref{t:minbaseswap} and Corollary~\ref{c:minmaxswap} imply a simple algorithm that allows to generate a superset of 
the non-dominated set for a given instance of Problem~\eqref{p:bbmp} by swapping between the optimal bases contained in $\b_i$ for $i=\{l,\ldots,u\}$. In this method, a sequence of minimal swaps has to be generated. The algorithm presented in~\cite{gabo:effi:84} uses a recursive procedure to generate this sequence. Further details on the generation of minimal swaps are given in Subsection~\ref{sec:bbmpgabo} below. 
Example~\ref{ex:bbmpswap} illustrates the idea of sequential minimal swaps at a graphic matroid.

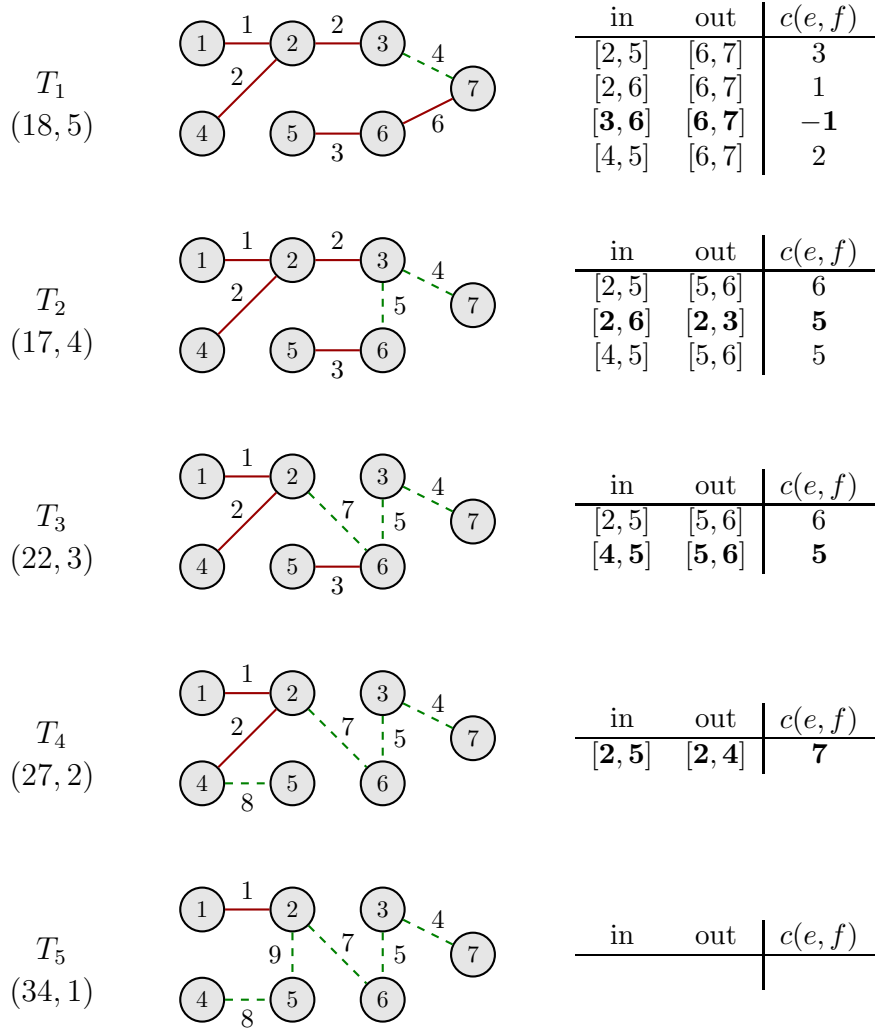
\begin{figure}[t]
\begin{center}
 \begin{tabular}{ccc}
\raisebox{-.25cm}{\begin{tabular}{c}\large$T_1$\\[0.1cm]\large$(18,5)$\end{tabular}}\hspace{0.5cm} &\raisebox{-1cm}{

\begin{tikzpicture}[scale=0.6]
\node[draw,circle,thick,fill=black!10, scale=0.9] (1) at (0,0)[] {\small 1};
\node[draw,circle,thick,fill=black!10, scale=0.9] (2) at (2,0) {\small 2};
\node[draw,circle,thick,fill=black!10, scale=0.9] (3) at (4,0) {\small 3};
\node[draw,circle,thick,fill=black!10, scale=0.9] (4) at (0,-2) {\small 4};
\node[draw,circle,thick,fill=black!10, scale=0.9] (5) at (2,-2) {\small 5};
\node[draw,circle,thick,fill=black!10, scale=0.9] (6) at (4,-2) {\small 6};
\node[draw,circle,thick,fill=black!10, scale=0.9] (7) at (6,-1) {\small 7};


\graph {
(1) --["\small 1",thick,draw=black!40!red] (2);
(2) --["\small 2",thick,draw=black!40!red] (3);
(2) --["\small 2",thick,swap,draw=black!40!red, inner sep=1pt] (4);
(5) --["\small 3",thick,swap,draw=black!40!red] (6);
(3) --["\small 4",thick,dashed,draw=black!50!green, inner sep=1pt] (7);
(6) --["\small 6",thick,swap,draw=black!40!red, inner sep=1pt] (7);
};
\end{tikzpicture}

}\hspace{0.5cm}  & \begin{tabular}{cc|c}
in 	& 	out 	& 	$c(e,f)$\\\hline
$[2,5]$ &	$[6,7]$	&	$3$\\
$[2,6]$ &	$[6,7]$	&	$1$\\
$\bf{[3,6]}$&	$\bf{[6,7]}$&	$\bf{-1}$\\
$[4,5]$ &	$[6,7]$	&	$2$
\end{tabular}
\\[1.55cm]
\raisebox{-.25cm}{\begin{tabular}{c}\large$T_2$\\[0.1cm]\large$(17,4)$\end{tabular}}\hspace{0.5cm} &\raisebox{-1cm}{

\begin{tikzpicture}[scale=0.6]
\node[draw,circle,thick,fill=black!10, scale=0.9] (1) at (0,0)[] {\small 1};
\node[draw,circle,thick,fill=black!10, scale=0.9] (2) at (2,0) {\small 2};
\node[draw,circle,thick,fill=black!10, scale=0.9] (3) at (4,0) {\small 3};
\node[draw,circle,thick,fill=black!10, scale=0.9] (4) at (0,-2) {\small 4};
\node[draw,circle,thick,fill=black!10, scale=0.9] (5) at (2,-2) {\small 5};
\node[draw,circle,thick,fill=black!10, scale=0.9] (6) at (4,-2) {\small 6};
\node[draw,circle,thick,fill=black!10, scale=0.9] (7) at (6,-1) {\small 7};


\graph {
(1) --["\small 1",thick,draw=black!40!red] (2);
(2) --["\small 2",thick,draw=black!40!red] (3);
(2) --["\small 2",thick,swap,draw=black!40!red, inner sep=1pt] (4);
(5) --["\small 3",thick,swap,draw=black!40!red] (6);
(3) --["\small 5",thick,dashed,draw=black!50!green] (6);
(3) --["\small 4",thick,dashed,draw=black!50!green, inner sep=1pt] (7);
};
\end{tikzpicture}

}\hspace{0.5cm}  & \begin{tabular}{cc|c}
in 	& 	out 	& 	$c(e,f)$\\\hline
$[2,5]$ &	$[5,6]$	&	$6$\\
$\bf{[2,6]}$&	$\bf{[2,3]}$&	$\bf{5}$\\
$[4,5]$ &	$[5,6]$	&	$5$
\end{tabular}
\\[1.55cm]
\raisebox{-.25cm}{\begin{tabular}{c}\large$T_3$\\[0.1cm]\large$(22,3)$\end{tabular}}\hspace{0.5cm} &\raisebox{-1cm}{

\begin{tikzpicture}[scale=0.6]
\node[draw,circle,thick,fill=black!10, scale=0.9] (1) at (0,0)[] {\small 1};
\node[draw,circle,thick,fill=black!10, scale=0.9] (2) at (2,0) {\small 2};
\node[draw,circle,thick,fill=black!10, scale=0.9] (3) at (4,0) {\small 3};
\node[draw,circle,thick,fill=black!10, scale=0.9] (4) at (0,-2) {\small 4};
\node[draw,circle,thick,fill=black!10, scale=0.9] (5) at (2,-2) {\small 5};
\node[draw,circle,thick,fill=black!10, scale=0.9] (6) at (4,-2) {\small 6};
\node[draw,circle,thick,fill=black!10, scale=0.9] (7) at (6,-1) {\small 7};


\graph {
(1) --["\small 1",thick,draw=black!40!red] (2);
(2) --["\small 2",thick,swap,draw=black!40!red, inner sep=1pt] (4);
(5) --["\small 3",thick,swap,draw=black!40!red] (6);
(2) --["\small 7",thick,dashed,draw=black!50!green, inner sep=1pt] (6);
(3) --["\small 5",thick,dashed,draw=black!50!green] (6);
(3) --["\small 4",thick,dashed,draw=black!50!green, inner sep=1pt] (7);
};
\end{tikzpicture}

}\hspace{0.5cm}  & \begin{tabular}{cc|c}
in 	& 	out 	& 	$c(e,f)$\\\hline
$[2,5]$ &	$[5,6]$	&	$6$\\
$\bf{[4,5]}$&	$\bf{[5,6]}$&	$\bf{5}$
\end{tabular}
\\[1.55cm]
\raisebox{-.25cm}{\begin{tabular}{c}\large$T_4$\\[0.1cm]\large$(27,2)$\end{tabular}}\hspace{0.5cm} &\raisebox{-1cm}{

\begin{tikzpicture}[scale=0.6]
\node[draw,circle,thick,fill=black!10, scale=0.9] (1) at (0,0)[] {\small 1};
\node[draw,circle,thick,fill=black!10, scale=0.9] (2) at (2,0) {\small 2};
\node[draw,circle,thick,fill=black!10, scale=0.9] (3) at (4,0) {\small 3};
\node[draw,circle,thick,fill=black!10, scale=0.9] (4) at (0,-2) {\small 4};
\node[draw,circle,thick,fill=black!10, scale=0.9] (5) at (2,-2) {\small 5};
\node[draw,circle,thick,fill=black!10, scale=0.9] (6) at (4,-2) {\small 6};
\node[draw,circle,thick,fill=black!10, scale=0.9] (7) at (6,-1) {\small 7};


\graph {
(1) --["\small 1",thick,draw=black!40!red] (2);
(2) --["\small 2",thick,swap,draw=black!40!red, inner sep=1pt] (4);
(4) --["\small 8",dashed,draw=black!50!green,thick,swap] (5);
(2) --["\small 7",thick,dashed,draw=black!50!green, inner sep=1pt] (6);
(3) --["\small 5",thick,dashed,draw=black!50!green] (6);
(3) --["\small 4",thick,dashed,draw=black!50!green, inner sep=1pt] (7);
};
\end{tikzpicture}

}\hspace{0.5cm}  & \begin{tabular}{cc|c}
in 	& 	out 	& 	$c(e,f)$\\\hline
$\bf{[2,5]}$ &	$\bf{[2,4]}$&	$\bf{7}$
\end{tabular}
\\[1.55cm]
\raisebox{-.25cm}{\begin{tabular}{c}\large$T_5$\\[0.1cm]\large$(34,1)$\end{tabular}}\hspace{0.5cm} &\raisebox{-1cm}{

\begin{tikzpicture}[scale=0.6]
\node[draw,circle,thick,fill=black!10, scale=0.9] (1) at (0,0)[] {\small 1};
\node[draw,circle,thick,fill=black!10, scale=0.9] (2) at (2,0) {\small 2};
\node[draw,circle,thick,fill=black!10, scale=0.9] (3) at (4,0) {\small 3};
\node[draw,circle,thick,fill=black!10, scale=0.9] (4) at (0,-2) {\small 4};
\node[draw,circle,thick,fill=black!10, scale=0.9] (5) at (2,-2) {\small 5};
\node[draw,circle,thick,fill=black!10, scale=0.9] (6) at (4,-2) {\small 6};
\node[draw,circle,thick,fill=black!10, scale=0.9] (7) at (6,-1) {\small 7};


\graph {
(1) --["\small 1",thick,draw=black!40!red] (2);
(4) --["\small 8",dashed,draw=black!50!green,thick,swap] (5);
(2) --["\small 9",dashed,thick,swap,draw=black!50!green] (5);
(2) --["\small 7",thick,dashed,draw=black!50!green, inner sep=1pt] (6);
(3) --["\small 5",thick,dashed,draw=black!50!green] (6);
(3) --["\small 4",thick,dashed,draw=black!50!green, inner sep=1pt] (7);
};
\end{tikzpicture}

}\hspace{0.5cm}  & \begin{tabular}{cc|c}
in 	& 	out 	& 	$c(e,f)$\\\hline
\phantom{$\bf{[2,5]}$} & \phantom{$\bf{[2,4]}$} &\phantom{$\bf{7}$}
\end{tabular}
\end{tabular}
\end{center}
\caption{Sequence of optimal spanning trees $\{T_1,\ldots,T_5\}$ for \eqref{p:bbmpeq} for the graphic matroid defined in Example~\ref{ex:bbmpswap}. Left column: tree $T_i$  and corresponding objective vector. Center column: associated tree. Right column: computation of a minimal swap $c(e,f)$ w.r.t.\ $T_i$, $i=1,\dots,5$.}\label{f:bbmpseq}
\end{figure}

\begin{example}\label{ex:bbmpswap}
We consider the graphic matroid induced by the graph $G=(V,E)$ given in Figure~\ref{f:bbmpex}. Note that $\X$ is the set of all spanning trees of $G$, and that the matroid has rank $m=6$. 
The objective coefficients of the first objective function $c$ are depicted next to each edge. For the second objective $b$, a solid red edge is used to indicate a cost of $1$, while a dashed green edge indicates a cost of $0$.

The spanning trees $T_1,\ldots,T_5$ given in Figure~\ref{f:bbmpseq} correspond to optimal solutions for Problem~\eqref{p:bbmpeq} for the right-hand side values $k\in\{1,\ldots,5\}$. We have that $T_i\in\b_i$, $i\in\{1,\ldots,5\}$ while $\b_0=\b_6=\emptyset$, i.e. $l=1$ and $u=5$. The objective vector $(c(T_i),b(T_i))$ of tree $T_i$, $i=1,\dots,5$, is stated in the first column, below the name of the respective tree. The corresponding trees are shown in the second column. The tables in the right-most column list relevant swaps w.r.t.\ the tree $T_i$, $i=1,\dots,5$, together with the respective cost, where minimal swaps are highlighted in bold. 
Here, the ``in''-column goes through the list of all dashed green edges that are not yet contained in $T_i$ and that may hence potentially be included. Adding the respective edges induces a unique cycle, and the best possible outgoing edge is shown in the ``out''-column. It is selected as a solid red edge in this cycle with maximum cost. 
Since we exchange a red against a green edge, the swap with minimal cost $c(e,f)$ w.r.t.\ $T_i$ leads to an optimal spanning tree $T_{i+1}\in\b_{i+1}$. While the spanning tree $T_1$ is dominated by $T_2$ (the implemented swap decreases each objective by one unit), the remaining trees form a complete set of efficient solutions and we conclude that $\Ynd=\{(17,4),(22,3),(27,2),(34,1)\}$.
\end{example}

Note that the procedure that is used to iteratively determine minimal swaps in Example~\ref{ex:bbmpswap} originates from \cite{gusf:matr:84}. In the following, we will present an improved procedure that avoids the computation of many unnecessary swaps. 
Example~\ref{ex:bbmpswap} further shows that not all optimal spanning trees for Problem~\eqref{p:bbmpeq} result in an efficient solution for Problem~\ref{p:bbmp}. However, we will show in the following that there exists a fixed index $j\in\{l,\ldots,u\}$ such that $B\in\b_i$ is efficient whenever $i\geq j$. Having a closer look at the example, it can be recognized that the minimal swaps that lead from $T_1$ to $T_5$ have non-decreasing 
costs. To prove that this property holds 
in general, we need the following lemma from \cite{gabo:effi:84}.

\begin{lemma}[see \cite{gabo:effi:84}, Lemma 3.2]\label{l:minswapb}
Let $B$ be a basis containing the element $e\in E_1 \cap B$. Let $(e,f)$ be a swap w.r.t.\ $B$ that has minimal cost among all swaps w.r.t.\ $B$ involving $e$, and set $B^\prime=B-e+f$. Given $g\in E_1\cap(B-e)$ arbitrary but fixed, let $(g,h)$ and $(g,h^\prime)$ denote swaps w.r.t.\ $B$ and $B^\prime$, respectively, that have minimal costs w.r.t.\ $B$ and $B'$, respectively, and that involve $g$. Then it holds that $c(g,h)\leq c(g,h^\prime)$.
\end{lemma}

Using Lemma~\ref{l:minswapb} it can now be shown that the sequence of costs induced by a sequence of minimal swaps is non-decreasing for increasing $i\in\{l,\ldots,u\}$.

\begin{theorem}\label{t:swapseqnondec}
Let $u\geq l+2$. For $i\in\{l,\ldots,u-1\}$ let $B_{i}\in\b_{i}$ and let $(e_i,f_i)$ denote a minimal swap w.r.t.\ $B_{i}$ leading to $B_{i+1}$. Then the sequence of costs of  minimal swaps $\{c(e_i,f_i)\}_{i=l}^{u-1}$ is non-decreasing, i.e. $c(e_{i},f_{i})\leq c(e_{i+1},f_{i+1})$ for all $i\in\{l,\ldots,u-2\}$.
\end{theorem}
\begin{proof}
Let $\{c(e_i,f_i)\}_{i=l}^{u-1}$ be a cost sequence of minimal swaps and let $i\in\{l,\ldots,u-2\}$ arbitrary but fixed. Note that $e_i\neq e_{i+1}$ since $e_i\in B_i\setminus B_{i+1}$. Moreover, $e_{i+1}\in B_i\cap B_{i+1}$ since otherwise $e_{i+1}$ would be contained in $B_{i+1}\setminus B_i$, i.e. $e_{i+1}=f_i$. But since $e_{i+1}$ is a red element of $\E$ while $f_i$ is a green element, this is impossible. 

Now consider a swap $(e_{i+1},f)$ w.r.t.\ $B_{i}$ that has minimal cost among all swaps w.r.t.\ $B_i$ that involve $e_{i+1}$. Note that the existence of a swap w.r.t.\ $B_i$ involving the edge $e_{i+1}$ follows from the basis exchange property \eqref{eq:B}: For the two bases $B_i,B_{i+2}$ we have $e_{i+1}\in B_i\setminus B_{i+2}$ and hence there exists an element $f\in B_{i+2}\setminus B_{i}$ such that $B_i-e_{i+1}+f\in\X$. Hence, $(e_{i+1},f)$ is a feasible swap w.r.t.\ $B_i$ involving $e_{i+1}$  and with $f\in B_{i+2}\setminus B_{i}=\{f_i,f_{i+1}\}$.

Since $(e_i,f_i)$ is a minimal swap w.r.t.\ $B_i$ it follows that $c(e_{i},f_{i})\leq c(e_{i+1},f)$. If $f=f_{i+1}$, we are done. Otherwise, we conclude from Lemma~\ref{l:minswapb} that $c(e_{i+1},f)\leq c(e_{i+1},f_{i+1})$, since the swap $(e_{i+1},f_{i+1})$ is minimal w.r.t.\ $B_{i+1}$. Combining these results we get $c(e_{i},f_{i})\leq c(e_{i+1},f_{i+1})$, which completes the proof.
\end{proof}

Since we have that
\begin{equation}\label{eq:baseswap}
 c(B_{i+1})-c(B_i)\,=\,c(f_i)-c(e_i)\,=\,c(e_i,f_i),
\end{equation}
Theorem~\ref{t:swapseqnondec} implies that the minimum costs of bases $B\in\b_i$ define a convex function for $i=|B\cap E_0|\in\{l,\ldots,u\}$.
Furthermore, if $c$ and $b$ are conflicting, then there must exist an index $j\in\{l,\ldots,u\}$ such that, starting from this index, all subsequent bases contained in the sequence $\{B_i\}_{i=j}^u$ correspond to  efficient solutions of Problem~\eqref{p:bbmp}. This holds since the value of the binary objective function $b$ is decreased by one unit when a swap from $B_i$ to $B_{i+1}$ is performed, while the corresponding value of the cost function $c$ remains constant or is increased. By construction, the index $j$ is the first index from $\{l,\ldots,u-1\}$ for which $c(e_{i},f_{i})> 0$ holds true. This implies the following result. 

\begin{theorem}\label{c:effbase}
Let $\{B_i\}_{i=l}^u$ denote the sequence of minimum cost bases such that $B_i\in\b_i$ for $i\in\{l,\ldots,u\}$. Assume that $u\geq l+2$. If there exists an index $j\in\{l+1,\ldots,u\}$ such that $c(B_{j-1})<c(B_{j})$, then $c(B_i)<c(B_{i+1})$ holds true for all $i\in\{j-1,\ldots,u-1\}$.
\end{theorem}
\begin{proof}
Let $j\in\{l+1,\ldots,u\}$ denote the index where $c(B_{j-1})<c(B_{j})$  holds true for the first time. If $j=u$ then there is nothing to show. So, let $j<u$. It suffices to prove that $c(B_j)<c(B_{j+1})$ holds true. From Equation~\eqref{eq:baseswap} it follows that $c(e_{j-1},f_{j-1})>0$. Furthermore, Theorem~\ref{t:swapseqnondec} implies that
\begin{equation*}
 c(B_{j+1})-c(B_j)\,=\,c(e_j,f_j)\,\geq\,c(e_{j-1},f_{j-1})\,>\,0,
\end{equation*} 
which shows that $c(B_j)<c(B_{j+1})$ is valid.
\end{proof}

Note that the basis $B_j$ where $j$ is the index such that $c(e_j,f_j)>0$ holds true for the first time is lexicographically optimal w.r.t.\ $c$ (with secondary optimization w.r.t.\ $b$). This means that $B_j$ is optimal w.r.t.\ $c$ and additionally satisfies $b(B_j)\leq b(B)$ for all $B\in\X$ with $c(B)=c(B_j)$. A lexicographically optimal basis $B_j$ can be computed efficiently using a greedy algorithm by computing an optimal basis  w.r.t.\ the costs $w(e)= (m+1) \cdot c(e)+b(e)$ for all $e\in E$, where $m$ is the rank of $\M$. 

Theorem~\ref{c:effbase} induces a method that generates a minimal complete set $\X_{cE}$ of efficient bases. Starting from a lexicographically optimal basis contained in $\b_j$, we compute a sequence of minimal swaps $\{(e_i,f_i)\}_{i=j}^{u-1}$ which is called \emph{swap sequence} in the following. By construction, we have that each of the generated bases $B_i$ is contained in $\b_i$ for $i\in\{j+1,\ldots,u\}$. The basis $B_j$ as well as all subsequently generated bases correspond to efficient solutions of Problem~\eqref{p:bbmp}. Note that starting with basis $B_j$ rather than with $B_l$ has the advantage that all generated bases are efficient. For example, for the graphic matroid from Example~\ref{ex:bbmpswap} the basis $B_l$ corresponds to $T_1$ in Figure~\ref{f:bbmpseq} while basis $B_j$ is given by $T_2$. Therefore, one unnecessary swap is omitted. Nevertheless, in the worst case $B_l=B_j$ holds and all swaps have to be calculated.

Since the binary objective $b$ decreases by one unit in each iteration of this procedure, it is ensured that no non-dominated outcome vector is missed in the objective space and hence $B_j,\ldots,B_u$ form a minimal complete set of efficient bases. Hence, we have proven the following result:

\begin{theorem}\label{c:bbmpnondomset}
Let $\{B_i\}_{i=j}^u$ denote a sequence of bases generated by a swap sequence. Then $X=\{B_j,\ldots,B_u\}$ forms a minimal complete set of efficient solutions and $\Ynd\,=\,\{(c(B_i),b(B_i)),\,i=j,\ldots,u\}$. 
\end{theorem}

\subsection{The Efficient Swap Algorithm}\label{sec:bbmpgabo}
The \emph{Efficient Swap Algorithm} (\esa) presented in this section utilizes swap sequences 
to efficiently generate a minimal complete set of efficient solutions for Problem~\eqref{p:bbmp}. Note that \esa\ can be interpreted as an extension of the algorithm stated in \cite{gabo:effi:84} for the solution of Problem~\eqref{p:bbmpeq} for fixed $k$.
Indeed, it was shown in \cite{gabo:effi:84} that this algorithm generates a complete swap sequence $\{(e_i,f_i)\}_{i=l}^{k-1}$ starting from $B_l\in\b_l$ and leading to $B_{k}\in\b_{k}$. Setting $k=u$ and starting from a lexicographically optimal basis $B_j$  thus induces \esa, and hence we omit detailed proofs for the correctness of this part of the algorithm. We rather focus on explaining how a complete swap sequence is generated without calculating a multiplicity of unnecessary swaps that do not lead to new efficient bases of the biobjective problem \eqref{p:bbmp}. At the end of this subsection we apply our algorithm to the graphic matroid from Example~\ref{ex:bbmpswap} to show how \esa\ works in practice.

We use the ideas from Theorem~\ref{t:minbaseswap} and Corollary~\ref{c:minmaxswap} to avoid the calculation of unnecessary swaps. 
In a first step, we generate bases $B_{j}\in\b_{j}$ and $B_u\in\b_u$ such that these two bases have as many elements as possible in common. 
The following two properties hold if and only if $B_{j}$ and $B_u$ coincide in a maximal number of elements:
\begin{itemize}
	\item[(a)] $B_{j}\cap E_0\subseteq B_u$, i.e., $B_u$ contains all green elements from $B_{j}$.
	\item[(b)] $B_u\cap E_1 \subseteq B_{j}$, i.e., $B_{j}$ contains all red elements from $B_u$.
\end{itemize}
Note that properties (a) and (b) imply that $U\coloneqq B_u\setminus B_{j}\subseteq E_0$ and that $\setJ\coloneqq B_{j}\setminus B_u\subseteq E_1$, respectively, and that $|U|=|\setJ|$.

If both properties (a) and (b) hold, then all elements of the matroid that are neither contained in $B_{j}$ nor in $B_u$ are redundant for \esa\ and can be removed from the ground set of the problem, i.e., we continue by considering the restriction $\M-(B_{j} \cup B_u)^c$. Furthermore, only those elements have to be swapped that are not contained in both bases simultaneously (see \cite{gabo:effi:84} for a detailed proof of this fact). This means that it is sufficient to consider the contraction of the matroid w.r.t.\ all elements that are contained in both bases. \esa\ works on this reduced problem $(\M-(B_{j} \cup B_u)^c)/(B_{j} \cap B_u)$ and uses a recursive swap sequence generation procedure ($\text{SSG}$) to generate a swap sequence. We will illustrate the main aspects of this procedure in the following, assuming that $B_{j}$ and $B_u$ satisfy properties (a) and (b) above and that we start from $B_{j}$.

If we add a green element $f$ from $B_u\setminus B_{j}\subseteq E_0$ with minimal costs to $B_{j}$, then a uniquely defined circuit $C(f,B_{j})$ is generated. Note that all elements of this circuit, with the only exception of $f$, are elements of $B_{j}$. If these elements are all red, then a minimal swap $(e^*,f)$ w.r.t.\ $B_{j}$ containing $f$, where $e^*\in C(f,B_{j})\setminus f$ and $c(e^*,f)\leq c(e,f)$ for all $e\in C(f,B_{j})\setminus f$, has to be contained in a swap sequence. The reason for this is that no other element of this circuit will lead to a better swap than the swap $(e,f)$ does, when $f$ is added to $B_{j}$. Otherwise, if the circuit $C(f,B_{j})$ contains red and green elements from $B_{j}$, 
then a minimal swap w.r.t.\ $B_{j}$ containing $f$ that is contained in a swap sequence cannot be deduced immediately. The idea in this case is to generate two smaller subproblems by contraction that do not intersect on the original ground set $\E$. The reduction to two subproblems is repeated until adding $f$ leads to a circuit with only red edges besides $f$.

As will be explained in the following, problem splitting can be realised such that all swaps that are already guaranteed to be contained in a final swap sequence by the criterion given above are preserved (see \cite{gabo:effi:84} for further details). Moreover, the problem can be split until adding $f$ leads to a circuit with only red edges besides $f$. Note that this is always satisfied when the respective ground sets of the contracted matroids consist of two elements $e\in B_{j}$ and $f\in B_u$ only. In this case, the swap $(e,f)$ must be contained in a final swap sequence since this swap is minimal.

\begin{algorithm}[t]
\caption{Efficient Swap Algorithm (\esa) for Biobjective Matroid Problems with one Binary Cost Function}\label{a:solvebbmp}
\begin{algorithmic}[1]
\INPUT An instance $((\M,\X,(c,b))$ of Problem~\eqref{p:bbmp}.
\OUTPUT $\Ynd$ and a complete set $\X_{cE}$ of efficient solutions. 
\STATE $\X_{cE}=\emptyset$, $\Ynd=\emptyset$.
\STATE Determine a lexicographically optimal basis $B_{j}$ 
\STATE Determine a minimum basis $B_u$ with respect to $c$ such that $B_u$ contains a maximal number of elements from $E_0$, all elements from $B_{j}\cap E_0$ and only those elements from $E_1$ that are also contained in $B_{j}$.
\STATE Call $\text{SSG}((\M-(B_{j} \cup B_u)^c)/(B_{j} \cap B_u),B_{j} \setminus B_u,B_u \setminus B_{j})$ to generate a swap sequence.
\STATE Let $\{(e_i,f_i)\}_{i={j}}^{u-1}$ denote the swap sequence found by Procedure $\text{SSG}$, where the swaps are sorted in non-decreasing order with respect to their costs.\label{a:bbmpsort2}
\STATE Set $B=B_{j}$, $\gamma=c(B_{j})$ and $\beta=b(B_{j})$.
\STATE $\X_{cE}=\{B\}$ and $\Ynd=\{(\gamma,\beta)\}$
\FOR {$i=j$ {\bf to} $u-1$ }
	\STATE Set $B=B-e_{i}+f_{i}$, $\gamma=\gamma+c(e_{i},f_{i})$ and $\beta=\beta-1$.
	\STATE Set $\X_{cE}=\X_{cE}\cup\{B\}$ and $\Ynd=\Ynd\cup\{(\gamma,\beta)\}$.
\ENDFOR
\RETURN $\X_{cE}$ and $\Ynd$.
\end{algorithmic}
\end{algorithm}

More formally, the split of the reduced matroid $(\M-(B_{j} \cup B_u)^c)/(B_{j} \cap B_u)$ into smaller parts is induced by a bisection of the sets $U=B_u\setminus B_{j}\subseteq E_0$ and $\setJ=B_{j}\setminus B_u\subseteq E_1$. At first, the sets $U$ and $\setJ$ are partitioned into two subsets $U_1,U_2$ and $\setJ_1,\setJ_2$ satisfying the following two conditions:
\begin{enumerate}
    \item The set $U_1\subseteq E_0$ consists of the $\lfloor|U|/2\rfloor$ smallest elements of $U$ with respect to~$c$.
    \item The set $B=\setJ_1\cup U_1$ is a minimum basis for $\M$ with respect to $c$ satisfying $B\cap E_0=U_1$.
\end{enumerate}
In a second step, the given problem is split into two different subproblems and the procedures $\text{SSG}((\M-U_2)/\setJ_1,\setJ_2,U_1)$ and $\text{SSG}((\M-\setJ_2)/U_1,\setJ_1,U_2)$ are executed. 

Applying this procedure, it can be shown (cf.~\cite{gabo:effi:84}) that all involved matroid problems remain feasible and that all swaps contained in the final swap sequence are preserved. Furthermore, if a subproblem consists of exactly one red element $e\in \setJ\subseteq (B_{j}\setminus B_u)\cap E_1$ and one green element $f\in U\subseteq (B_u\setminus B_{j})\cap E_0$, it is guaranteed that the swap $(e,f)$ is in the  swap sequence.

\begin{algorithm}[t]
\caption{Swap Sequence Generation $\text{SSG}(\M,\setJ,U)$ (\cite{gabo:effi:84})}\label{a:procbbmp}
\begin{algorithmic}[1]
\INPUT A matroid $\M$ and two sets of elements $\setJ\subseteq B_{j}\setminus B_u$ and $U\subseteq B_u\setminus B_{j}$, $|\setJ|=|U|$.
\OUTPUT A minimal swap $(e,f)$ or two recursive calls of the procedure SSG.
\IF {$|U|=1$}
	\RETURN the swap $(e,f)$, where $\setJ=\{e\}$ and $U=\{f\}$.
\ELSE
	\STATE Let $U_1$ be the set of $\lfloor|U|/2\rfloor$ smallest 
	elements with respect to $c$ (contained in $E_0$) and set $U_2=U\setminus U_1$.\label{a:bbmpsort1}
	\STATE Determine $\setJ_1$ such that $B=\setJ_1\cup U_1$ forms a minimal basis for $\M$ with respect to $c$ satisfying $B\cap E_0=U_1$ and set $\setJ_2=\setJ\setminus \setJ_1$.
	\STATE Call $\text{SSG}((\M-U_2)/\setJ_1,\setJ_2,U_1)$ to find the swaps for the elements in $U_1$.
	\STATE Call $\text{SSG}((\M-\setJ_2)/U_1,\setJ_1,U_2)$ to find the swaps for the elements in $U_2$.
\ENDIF
\end{algorithmic}
\end{algorithm}

The Efficient Swap Algorithm {\esa} for the solution of Problem~\eqref{p:bbmp} is summarized in Algorithm~\ref{a:solvebbmp}. The associated bisection procedure $\text{SSG}$ that is recursively called during the course of {\esa} is outlined in Algorithm~\ref{a:procbbmp}. 
At the beginning of Algorithm~\ref{a:solvebbmp} the two  bases $B_{j}$ and $B_u$ are calculated. Then, using Algorithm~\ref{a:procbbmp}, a swap sequence for the (contracted) matroid $(\M-(B_{j} \cup B_u)^c)/(B_{j} \cap B_u))$ with ground set $(B_{j} \cup B_u)\setminus(B_{j} \cap B_u)$ is generated recursively. Finally, the generated swaps are sorted in non-decreasing order of their costs and, based on the result of Theorem~\ref{c:bbmpnondomset}, the  non-dominated set $\Ynd$ as well as a minimal complete set $\X_{cE}$ of efficient solutions are determined.

Note that during the course of Algorithm~\ref{a:procbbmp} it may happen that swaps (or elements) with the same cost occur. So, a rule how to cope with ties 
in Line~\ref{a:bbmpsort1} of Algorithm~\ref{a:procbbmp} has to be given. We follow the approach suggested in \cite{gabo:effi:84}: 
First assume that the elements of $E_0$ are sorted and indexed according to their costs $c$ in non-decreasing order. Then, in Line~\ref{a:bbmpsort1} of Algorithm~\ref{a:procbbmp} we always choose the  first $\lfloor|U|/2\rfloor$ elements from $U$. When there are ties in the costs of the swap sequence, 
then the affected swaps are arranged in increasing order of the indices with respect to the elements that are contained in $E_0$. 
The following theorem summarizes the results.

\begin{theorem}
 Algorithm~\ref{a:solvebbmp} is correct and returns the non-dominated set and a minimal complete set of efficient solutions.
\end{theorem}
\begin{proof}
 The correctness of the algorithm follows from Theorem~\ref{c:bbmpnondomset} and from the correctness of the algorithm for solving Problem~\eqref{p:bbmpeq} stated in \cite{gabo:effi:84}.
\end{proof}

Note that the complexity of Algorithm~\ref{a:solvebbmp} depends on the considered matroid problem. For graphic matroids with $G=(V,E)$, for example, it is shown in~\cite{gabo:effi:84} that their basic algorithm solves Problem~\eqref{p:bbmpeq} within $\bigo(m\log \log_{(2+m/n)}n+n\cdot\log(n))$ time, where $|V|=n$ and $|E|=m$. Hence,  Algorithm~\ref{a:solvebbmp} has the same time bound in this case, since the additional construction of $\X_{cE}$ and $\Ynd$ takes at most $\bigo(m)$ time. 
For a matching matroid and a transversal matroid the time bound is $\bigo(n\log n+m\ell)$, which follows again from a corresponding result in \cite{gabo:effi:84}, where $n$ is the number of vertices of a graph, $\ell$ is the number of edges in a maximum matching and $m$ is the number of edges in the graph. Again, Algorithm~\ref{a:solvebbmp} has the same time bound, since the additional construction of $\X_{cE}$ and $\Ynd$ takes at most $\bigo(\ell)$ time. Furthermore, it is proven in \cite{gabo:effi:84} that the Problem~\eqref{p:bbmpeq} can be solved in linear time, i.e.\ $\bigo(n)$, for a partition matroid (and therefore also for a uniform matroid) on a groundset $\E$ which consists of $n$ elements. In this case the construction of $\X_{cE}$ and $\Ynd$ takes at most $\bigo(n)$ time and hence Algorithm~\ref{a:solvebbmp} has the same time bound.

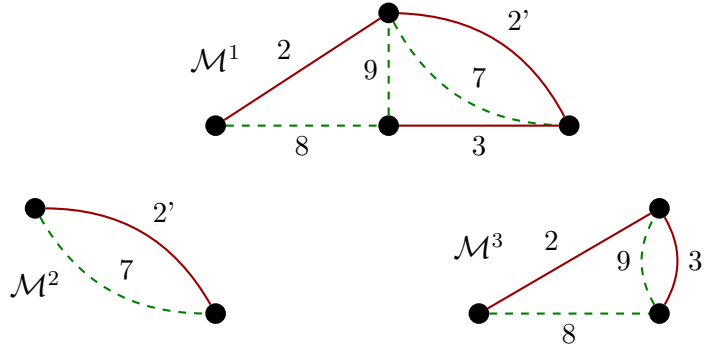
\begin{figure}[t]
\begin{center}

\begin{tikzpicture}[scale=0.1]
\node[draw,circle,thick,fill=black,inner sep=0pt,minimum size=7pt] (1) at (25,26)[] {};
\node[draw,circle,thick,fill=black,inner sep=0pt,minimum size=7pt] (2) at (48,26) {};
\node[draw,circle,thick,fill=black,inner sep=0pt,minimum size=7pt] (3) at (48,41) {};
\node[draw,circle,thick,fill=black,inner sep=0pt,minimum size=7pt] (5) at (72,26) {};

\node[draw,circle,thick,fill=black,inner sep=0pt,minimum size=7pt] (6) at (1,15) {};
\node[draw,circle,thick,fill=black,inner sep=0pt,minimum size=7pt] (8) at (25,1) {};

\node[draw,circle,thick,fill=black,inner sep=0pt,minimum size=7pt] (9) at (60,1) {};
\node[draw,circle,thick,fill=black,inner sep=0pt,minimum size=7pt] (10) at (84,1) {};
\node[draw,circle,thick,fill=black,inner sep=0pt,minimum size=7pt] (11) at (84,15) {};

\node (A) at (25,35) [circle] {\large{$\M^1$}};
\node (B) at (1,5) [circle] {\large{$\M^2$}};
\node (C) at (60,10) [circle] {\large{$\M^3$}};


\graph {
(1) --["2",thick,draw=black!40!red] (3);
(1) --["8",thick,swap,dashed,draw=black!50!green] (2);
(2) --["9",thick,dashed,draw=black!50!green] (3);
(3) --["7",thick,dashed,draw=black!50!green, bend right] (5);
(2) --["3",thick,swap,draw=black!40!red] (5);
(5) --["2'",thick,swap,draw=black!40!red, bend right] (3);

(6) --["2'",thick,draw=black!40!red, bend left] (8);
(6) --["7",thick,dashed,draw=black!50!green, bend right] (8);

(9) --["8",thick,swap,dashed,draw=black!50!green] (10);
(9) --["2",thick,draw=black!40!red] (11);
(10) --["9",thick,dashed,draw=black!50!green,bend left] (11);
(10) --["3",thick,swap,draw=black!40!red,bend right] (11);
};
\end{tikzpicture}

\end{center}
\caption{Contracted graphic matroids $\M^1$, $\M^2$ and $\M^3$ 
from Example~\ref{ex:bbmpcont}. Solid red lines correspond to edges $e$ with $b(e)=1$ while the green dashed lines correspond to edges with $b(e)=0$.  The edges are identified by their associated cost value~$c$, where ambiguities are resolved by using the notation $2$ and $2^\prime$ to refer to the edges $[2,4]$ and $[2,3]$, respectively.}\label{f:bbmpshrink}
\end{figure}

\begin{example}\label{ex:bbmpcont}
We apply {\esa} to the graphic matroid introduced in Example~\ref{ex:bbmpswap}. 
To simplify the notation, the edges of the graph $G$ (see Figure~\ref{f:bbmpex}) are identified by their associated costs $c$ rather than by their respective end nodes. This only induces ambiguity in the case of the edges $[2,3]$ and $[2,4]$ which both have cost $2$, and in the case of the edges $[1,4]$ and $[3,7]$ which both have cost $4$. We will refer to the edge $[2,3]$ by writing $2^\prime$ and to the edge $[1,4]$ by writing $4^\prime$ in the following to distinguish between these edges. 

In a first step the optimal bases $B_{j}$ and $B_u$ are determined. This leads to the spanning trees $T_{2}$ and $T_5$, respectively, shown in Figure~\ref{f:bbmpseq}, i.e.\  $B_{j}=\{1,2,2^\prime,3,4,5\}$ and $B_u=\{1,4,5,7,8,9\}$. 
Hence, $U=B_u\setminus B_{j}=\{7,8,9\}\subseteq E_0$, $\setJ=B_{j}\setminus B_u=\{2,2^\prime,3\}\subseteq E_1$, and $B_{j}\cap B_u=\{1,4,5\}$. This implies that the edges $1$, $4$ and $5$, i.e., the edges $[1,2]$, $[3,6]$ and $[3,7]$, are contained in every efficient spanning tree in the set $\X_{cE}$ generated by \esa, and the edges $4^\prime$ and $6$, i.e., the edges $[1,4]$ and $[6,7]$, can be removed from the problem since they are not contained in $B_{j}\cup B_u$. The contracted matroid $\M^1\coloneqq (\M-(B_{j} \cup B_u)^c)/(B_{j} \cap B_u)$ is shown in Figure~\ref{f:bbmpshrink}. From now on, we will enumerate (contracted) matroids and their respective subsets by superscripts, while referring to the corresponding subsets $U_1,U_2,J_1,J_2$ by subscripts, as before.

Then the procedure $\text{SSG}$ is called with $\text{SSG}(\M^1,\setJ^1,U^1)$, where $\setJ^1\coloneqq \setJ=\{2,2^\prime,3\}$ and $U^1\coloneqq U=\{7,8,9\}$. Since $|U^1|=3>1$, the matroid $\M^1$ has to be split into two smaller matroids. We first determine the $\lfloor|U^{1}|/2\rfloor$ smallest elements of $U^{1}$ as $U_1^1=\{7\}$  and set $U_2^1\coloneqq U^1\setminus U^1_1=\{8,9\}$. Now we determine $\setJ_1^1$ such that $B^1=\setJ_1^1\cup U^1_1$ is a minimum basis for $\M^1$ with respect to $c$ satisfying $B^1\cap E_0=U^1_1$. This implies that $\setJ_1^1=\{2,3\}$, $\setJ_2^1\coloneqq \setJ^1\setminus \setJ_1^1=\{2^\prime\}$ and $B^1=\{2,3,7\}$. 
Now the procedure $\text{SSG}$ is called recursively with $\text{SSG}(\M^2,\setJ^2,U^2)$ and $\text{SSG}(\M^3,\setJ^3,U^3)$, respectively, where $\M^2$ and $\M^3$ correspond to the contracted matroids shown in Figure~\ref{f:bbmpshrink}, and $\setJ^2=\setJ_2^1$, $\setJ^3=\setJ_1^1$, $U^2=U^1_1$ and $U^3\coloneqq U^1\setminus U_1^1=\{8,9\}$. $\text{SSG}(\M^2,\setJ^2,U^2)$ returns immediately the swap $(2^\prime,7)$ while $\text{SSG}(\M^3,\setJ^3,U^3)$ needs another recursion to compute the swaps $(3,8)$ and $(2,9)$. 
Sorting these swaps in non-decreasing order of their costs leads to the swap sequence $\{(2^\prime,7),(3,8),(2,9)\}$ with costs $5,5,7$. This immediately leads to the final result 
$\Ynd=\{(17,4),(22,3),(27,2),(34,1)\}$  and $\X_{cE}=\{T_2,T_3,T_4,T_5\}$, see also Figure~\ref{f:bbmpseq}.
\end{example}

\section{Connectedness of the Efficient Set}\label{s:basescon}

In the following we show that the set of efficient bases $\Xeff$ for Problem~\eqref{p:bbmp} is always connected. We recall from Section~\ref{subsec:matroids} that the set $\Xeff$ is said to be connected if its corresponding adjacency graph is connected. Recall also that two efficient bases of a matroid of rank $m$ are called adjacent if they have $m-1$ elements in common. Our proof is based on the fact that the set of \emph{supported} efficient bases is always connected with respect to the above given definition of adjacency for efficient bases. For more details on this topic we refer to~\cite{ehrg:onma:96}. In the following we show that every efficient basis of Problem~\eqref{p:bbmp} is a supported efficient solution which implies that the adjacency graph of the problem is always connected.

To do so, we first formulate a sufficient condition that guarantees that the non-dominated set of a general biobjective combinatorial minimization problem only consists of supported non-dominated outcome vectors. Given the non-dominated set $\Ynd=\{z_1,\ldots,z_n\}\subset\R^2$ of the problem, where $n\geq 3$ and $z_i=(x_i,y_i)\in\R^2$, with $x_1<\ldots<x_n$ and $y_1>\ldots>y_n$, we define the \emph{sequence of slopes} $\{m_i\}_{i=1}^{n-1}$ of subsequent points of $\Ynd$ by setting
\[m_i\,=\,\dfrac{y_{i+1}-y_{i}}{x_{i+1}-x_{i}},\quad i=1,\dots,n-1.\]
Note that $m_i\in(-\infty,0)$ holds for all $i\in\{1,\ldots,n-1\}$.

\begin{lemma}\label{t:slope} Consider a  biobjective combinatorial minimization problem and suppose that the sequence of slopes $\{m_i\}_{i=1}^{n-1}$ is non-decreasing. Then all non-dominated outcome vectors in the set $\Ynd$ are supported.
\end{lemma}
\begin{proof}
Suppose that, to the contrary, there is a non-supported non-dominated outcome vector $z_t\in\Ynd$, $t\in\{2,\ldots,n-1\}$. Since a non-dominated outcome vector is supported if and only if it is an element of the convex hull of $\Y$, it follows that there exist supported non-dominated outcome vectors $z_i,z_j\in\Ynd$ and a weight  $\lambda\in(0,1)$ such that the point $z_\lambda=(x_\lambda,y_\lambda)\coloneqq \lambda z_i+(1-\lambda) z_j\in\R^2$ strongly dominates $z_t$, where $1\leq i<t<j\leq n$ holds. Note that $z_\lambda$ can not be an element of $\Ynd$ since otherwise it would dominate $z_t$. Without loss of generality we may assume that $i=1$ and $t=2$.
Since $x_1<x_\lambda<x_2$ and $y_\lambda<y_2<y_1$ holds, it follows that
\[ (y_\lambda-y_1)\cdot(x_2-x_1)\,<\,(y_2-y_1)\cdot(x_2-x_1)\,<\,(y_2-y_1)\cdot(x_\lambda-x_1)\,<\,0.\]
Since $z_\lambda$ is an element of the straight line connecting $z_1$ and $z_j$, it follows that
\[
m^\star\coloneqq \dfrac{y_j-y_1}{x_j-x_1}\,=\,\dfrac{y_\lambda-y_1}{x_\lambda-x_1}\,<\,\dfrac{y_2-y_1}{x_2-x_1}\,=\,m_1.
\]
This is impossible, since by assumption $m_1\leq m_i$ for all $i\in\{1,\ldots,n\}$, and hence
\begin{eqnarray*}
 y_j	&=&y_1+\sum_{i=1}^{j-1}(y_{i+1}-y_i)\,=\,y_1+\sum_{i=1}^{j-1}m_i\cdot(x_{i+1}-x_i)\\
	&\geq&y_1+m_1\cdot\sum_{i=1}^{j-1}(x_{i+1}-x_i)\,=\,y_1+m_1\cdot(x_j-x_1).
\end{eqnarray*}
Therefore, it has to hold that $m^\star\geq m_1$, which is a contradiction.
\end{proof}

We combine the results of Theorem~\ref{c:effbase}, Theorem~\ref{c:bbmpnondomset} and Lemma~\ref{t:slope} to conclude that all nondominated outcome vectors of biobjective optimization problems on matroids with one binary objective function are supported.
\begin{lemma}\label{t:bstpycon}
Consider a feasible instance of Problem~\eqref{p:bbmp}, i.e., assume that $\Y\neq\emptyset$. Then the non-dominated set $\Ynd$ consists only of supported non-dominated outcome vectors.
\end{lemma}
\begin{proof}
Using the notation introduced in Section~\ref{sec:solvebbmp}, we denote by $\{(e_i,f_i)\}_{i=j}^{u-1}$ a swap sequence starting from a lexicographically optimal basis $B_j$ that induces a set of efficient bases $B_i\in\b_i$ 
for Problem \eqref{p:bbmp}, $i=j,\ldots,u$. 
According to Theorem~\ref{c:bbmpnondomset} we have that $\Ynd=\{(c(B_i),b(B_i)),\,i=j,\ldots,u\}$. 

Note that the result is trivial when $|\Ynd|\leq 2$. When $|\Ynd|\geq 3$ we know from Lemma~\ref{t:slope} that it suffices to show that the sequence of slopes $\{m_i\}_{i=j}^{u-1}$, where
\[
m_i\,=\,\dfrac{b(B_{i+1})-b(B_{i})}{c(B_{i+1})-c(B_{i})}\,=\,\dfrac{-1}{c(B_{i+1})-c(B_{i})}
\]
is non-decreasing.
Since in this case $|\Ynd|\geq 3$ we have that $j\leq u-2$. For an arbitrary but fixed index $i\in\{j,\ldots,u-2\}$ it follows from Theorem~\ref{t:swapseqnondec} and Theorem~\ref{c:effbase} that
\begin{equation*}
 c(B_{i+2})-c(B_{i+1})\,=\,c(e_{i+1},f_{i+1})\,\geq\,c(e_{i},f_{i})\,=\,c(B_{i+1})-c(B_i)\,>\,0.
\end{equation*}
This implies that
\[
m_{i+1}\,=\,\dfrac{-1}{c(B_{i+2})-c(B_{i+1})}\,\geq\,\dfrac{-1}{c(B_{i+1})-c(B_{i})}\,=\,m_i,
\]
and hence the sequence of slopes $\{m_i\}_{i=j}^{u-1}$ is non-decreasing. This implies that $\Ynd$ contains only supported non-dominated outcome vectors.
\end{proof}
\begin{figure}[tbh]
\begin{center}
  \begin{tikzpicture}[scale=0.4]
\draw[very thin,color=gray] (16,0) grid (39,6);
\fill (17,4) circle (2ex);
\node (A) at (16.5,3.2) [circle] {$y^2$};
\fill (22,3) circle (2ex);
\node (B) at (21.5,2.2) [circle] {$y^3$};
\fill (27,2) circle (2ex);
\node (C) at (26.5,1.2) [circle] {$y^4$};
\fill (34,1) circle (2ex);
\node (D) at (33.2,0.5) [circle] {$y^5$};
\node (E) at (17.5,5.5) [circle] {$y^1$};
\draw[dotted] (18,5) to (17,4);
\draw[-] (17,4) to (22,3);
\draw[-] (22,3) to (27,2);
\draw[-] (27,2) to (34,1);

\draw[fill=black!30] (37,1) circle (2ex);
\draw[fill=white] (25,3) circle (2ex);
\draw[fill=white] (27,4) circle (2ex);
\draw[fill=white] (28,4) circle (2ex);
\draw[fill=white] (35,3) circle (2ex);
\draw[fill=white] (33,3) circle (2ex);
\draw[fill=black!30] (31,2) circle (2ex);
\draw[fill=black!30] (26,3) circle (2ex);
\draw[fill=black!30] (33,2) circle (2ex);
\draw[fill=black!30] (23,3) circle (2ex);
\draw[fill=black!30] (30,2) circle (2ex);
\draw[fill=black!30] (28,2) circle (2ex);
\draw[fill=white] (30,3) circle (2ex);
\draw[fill=white] (25,4) circle (2ex);
\draw[fill=white] (32,3) circle (2ex);
\draw[fill=white] (21,5) circle (2ex);
\draw[fill=white] (19,5) circle (2ex);
\draw[fill=black!30] (25,3) circle (2ex);
\draw[fill=white] (26,4) circle (2ex);
\draw[fill=white] (24,4) circle (2ex);
\draw[fill=black!30] (19,4) circle (2ex);
\draw[fill=black!30] (24,3) circle (2ex);
\draw[fill=black!30] (27,3) circle (2ex);
\draw[fill=black!30] (20,4) circle (2ex);
\draw[fill=white] (29,4) circle (2ex);
\draw[fill=white] (22,5) circle (2ex);
\draw[fill=white] (29,3) circle (2ex);
\draw[fill=white] (31,3) circle (2ex);
\draw[fill=white] (36,2) circle (2ex);
\draw[fill=black!30] (29,2) circle (2ex);
\draw[fill=black!30] (32,2) circle (2ex);
\draw[fill=white] (39,2) circle (2ex);
\draw[fill=white] (34,3) circle (2ex);
\draw[fill=white] (32,4) circle (2ex);
\draw[fill=white] (36,3) circle (2ex);
\draw[fill=white] (38,2) circle (2ex);
\draw[fill=black!30] (34,2) circle (2ex);
\draw[fill=white] (35,2) circle (2ex);
\draw[fill=white] (24,5) circle (2ex);
\draw[fill=white] (23,4) circle (2ex);
\draw[fill=white] (28,3) circle (2ex);
\draw[fill=black!30] (22,4) circle (2ex);
\draw[fill=white] (30,4) circle (2ex);
\draw[fill=white] (25,5) circle (2ex);
\draw[fill=white] (26,5) circle (2ex);
\draw[fill=white] (20,5) circle (2ex);
\draw[fill=white] (23,5) circle (2ex);
\draw[fill=black!30] (21,4) circle (2ex);
\draw[fill=white] (18,5) circle (2ex);

\draw[->,thick] (16,0) -- (40,0) node[right] {$c$};
\draw[->,thick] (16,0) -- (16,7) node[above] {$b$};
\foreach \x in {16,17,18,19,20,21,22,23,24,25,26,27,28,29,30,31,32,33,34,35,36,37,38,39}\draw (\x,-0.1) -- (\x,0.1) node[below=5pt] {$\scriptstyle\x$};
\foreach \y in {0,1,2,3,4,5,6}\draw (15.9,\y) -- (16.1,\y) node[left=5pt]  {$\scriptstyle\y$};
\end{tikzpicture} 
\end{center}
\caption{Non-dominated set of the graphic matroid introduced in Figure~\ref{f:bbmpex}. Non-dominated outcome vectors are shown in black and weakly non-dominated points shown in grey. The remaining (white) points are all dominated outcome vectors. Note that the points $y^2$, $y^4$ and $y^5$ are extreme supported non-dominated outcome vectors while $y^3$ is a non-extreme supported non-dominated outcome vector. The nodes $y^i$ correspond to the trees $T_i$ from Figure~\ref{f:bbmpseq}.} \label{f:zfktraum}
\end{figure}
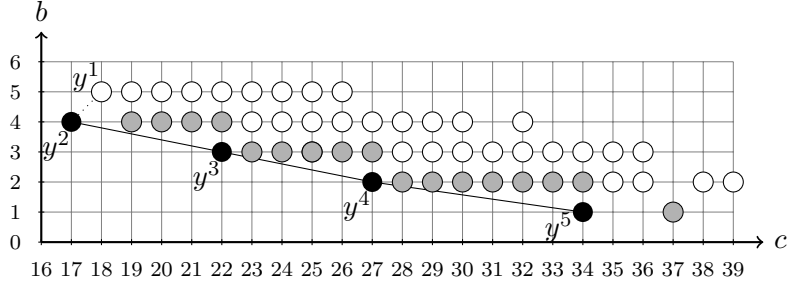
Note that not every supported non-dominated outcome vector must be extreme supported. Indeed, when the costs of two consecutive swaps $(e_i,f_i)$ and $(e_{i+1},f_{i+1})$ in a swap sequence are equal, then the point $(c(B_{i+1}),b(B_{i+1}))$ is not an extreme point of $\conv(\Y)$. To see this we consider again the graphic matroid introduced in Figure~\ref{f:bbmpex}, c.f.\  Example~\ref{ex:bbmpswap}. The set of feasible outcome vectors $\Y$ in the objective space is shown in Figure~\ref{f:zfktraum}. In this example, the supported non-dominated point $y^3$ (which is the image of the spanning tree $T_3$ from Figure~\ref{f:bbmpseq}) is not an extreme point of $\conv(\Y)$ and thus not extreme supported.

We finally conclude that the set of efficient bases is connected.

\begin{theorem}\label{cor:bbmpeffcon}
Consider a feasible instance of Problem~\eqref{p:bbmp}. Then the set of efficient solutions $\Xeff$ is connected.
\end{theorem}
\begin{proof}
Lemma~\ref{t:bstpycon} implies that all non-dominated outcome vectors in $\Y_N$ are supported, and hence all efficient solutions in $\X_E$ are supported. Using the fact that the (sub)graph of all supported efficient solutions is always connected (see \cite{ehrg:onma:96}) implies the result.
 \end{proof}
 
 \section{Numerical Results}\label{sec:bbmpnumres}
 The main advantage of \esa\ is its computational  efficiency. In this section we present numerical results that validate this statement for the examples of graphic and uniform matroids.
In addition we address the question whether, and if yes, how far the results on the connectedness of the efficient set can be extended to more general cases. Towards this end, we randomly generated instances of uniform matroids with more than two (integer) values for the coefficients in the second objective. For all 
 instances we compute the complete efficient set $\Xeff$ and count the number of instances for which $\Xeff$ is non-connected.

 We note that other generalizations have been investigated for specific matroids. \cite{seipp:diss:2013}, for example, analyzes uniform matroids with one general cost function and two binary cost functions in a tri-criteria model. They suggest an exact solution method that is, similar to \esa, based on neighborhood search. In contrast to \esa\ their algorithm may generate dominated solutions. Nevertheless, they show that a complete set of efficient solutions can be generated in polynomial time with this method and that the efficient set consists only of supported solutions and is thus connected.
 
 \subsection{Performance of the Efficient Swap Algorithm}\label{subsec:numESA}
 
 In this section, we present numerical results on randomly generated instances of graphic matroids and of uniform matroids to validate the efficiency of \esa.
 
 \subsubsection{Graphic Matroids}
For graphic matroids on undirected connected graphs $G=(V,E)$, i.e., for biobjective  minimum spanning tree problems with one general and one binary cost function, we evaluate the computational time needed by \esa\ to compute the non-dominated set $\Ynd$. To set this time in relation to the combinatorial complexity of the respective instances, we also provide the total number of feasible solutions, i.e., of spanning trees of the graph, and evaluate the time needed to determine all efficient trees from this set by total enumeration. This complete enumeration approach (\ce) is implemented by using the matlab code by Matthias Hotz \cite{hotz:matlab:2020} for the generation of all spanning trees that is based on an algorithm 
described in \cite{knuth:2012}. For a recent survey and numerical comparison of exact algorithms for general multiobjective minimum spanning tree problems we refer to \cite{fern:empi:2020}.
Note that the problem could also be solved by $n=|V|$ restarts of the method of Gabow and Tarjan \cite{gabo:effi:84} with appropriately chosen constraints on the number of green edges. \esa\ avoids these restarts as well as the computation of dominated solutions by initializing the swap sequence with a lexicographically optimal basis. The induced savings depend on the considered instance and are most significant 
when the non-dominated set is rather small compared to $|V|$.
 
The efficiency tests are run on a computer with an Intel(R) Core(TM) i7-8700 CPU @ 3.20GHz processor, 12MB Cache and 32 GB RAM. Both algorithms are implemented in MATLAB Version R2020a.
 
Recall from Section~\ref{sec:bbmp} that the objective values of the binary objective $b$ can only take values between $0$ and $m$, i.e., $b(B)\in\{0,1,\dots,m\}$ where $m$ is the rank of the underlying matroid and $B$ is an arbitrary basis. As a consequence, we have that  $|\Ynd|=\bigo(m)$. For an instance of the graphic matroid on a connected graph $G=(V,E)$ with $n$ vertices and $m$ edges, this implies that $b(T)\in\{0,1,\dots,n-1\}$ for all spanning trees $T$ of $G$. Note that due to the common notation that $|V|=n$ and $|E|=m$ for graphic matroids, the rank of a graphic matroid is thus $n-1$.  
 The \ce\ approach determines all efficient spanning trees by maintaining a list with $n$ entries, one for each potential value of $b(T)\in\{0,1,\dots,n-1\}$ that stores the currently best cost value $c(T)$ together with all corresponding trees that were enumerated so far.

 \begin{table}
     \centering
     \tbl{Computational results for randomly generated graphs with $n$ vertices, $m$ edges, and $\vert E_1\vert$ edges with cost $1$ in the binary cost function $b$. $|\Xeff|$ is the number of efficient solutions, $|\Ynd|$ is the number of non-dominated points and $|\X|$ is the number of spanning trees for this instance. 
     The last two columns give the 
     time in seconds for \esa\ and for \ce, respectively.}
{\begin{tabular}{c|r|r|r|l|r|r}
($n$, $m$) & $\vert E_1\vert$ &$|\Ynd|$ & $|\Xeff|$ & $|\X|$ & \esa\ [s] & \ce\ [s] \\
\hline
($7$, $10$) & $2$ &$1$ &$1$ &\multicolumn{1}{r|}{$76$} &$0.065$&$0.010$\\
 ($7$, $10$) & $4$ &$4$&$4$ &\multicolumn{1}{r|}{$66$} &$0.032$&$0.001$\\
 ($7$, $15$) & $8$ &$2$ &$2$&\multicolumn{1}{r|}{$1\,615$} &$0.010$&$0.013$\\
  ($7$, $15$) & $12$ &$3$ &$3$&\multicolumn{1}{r|}{$1\,807$} &$0.014$&$0.015$\\
 ($7$, $20$) & $8$ &$4$&$4$&\multicolumn{1}{r|}{$12\,005$} &$0.019$&$0.080$\\
 ($7$, $20$) & $11$ &$5$ &$5$ &\multicolumn{1}{r|}{$12\,005$} &$0.023$&$0.081$\\
 ($7$, $20$) & $12$ &$5$ &$5$ &\multicolumn{1}{r|}{$12\,005$} &$0.022$&$0.081$\\
 ($10$, $20$) & $10$ &$6$ &$6$ &\multicolumn{1}{r|}{$26\,646$} &$0.027$&$0.203$\\
 ($10$, $20$) & $11$ &$5$ &$5$ &\multicolumn{1}{r|}{$21\,560$} &$0.023$&$0.167$\\
 ($10$, $20$) & $15$ &$2$ &$2$ &\multicolumn{1}{r|}{$18\,956$} &$0.011$&$0.139$\\
 ($10$, $30$) & $15$ &$3$ &$3$ &$1.85\cdot 10^6$ &$0.016$&$11.808$\\
 ($10$, $30$) & $17$ &$2$ &$2$&$1.62\cdot 10^6$ &$0.012$&$10.215$\\
 ($10$, $30$) & $20$ &$5$ &$5$&$1.60\cdot 10^6$ &$0.021$&$10.202$\\
 ($10$, $40$) & $17$ &$6$ &$6$&$3.06\cdot 10^7$ &$0.029$&$172.213$\\
 ($10$, $40$) & $18$ &$7$ &$7$&$3.01\cdot 10^7$ &$0.032$&$171.100$\\
 ($10$, $40$) & $20$ &$3$ &$3$&$3.01\cdot 10^7$ &$0.016$&$168.403$\\
 ($15$, $30$) & $13$ &$5$ &$5$ &$5.35\cdot 10^6$ &$0.025$&$38.684$\\
 ($15$, $30$) & $15$ &$5$ &$5$ &$4.11\cdot 10^6$ &$0.023$&$29.293$\\
 ($15$, $30$) & $16$ &$5$ &$5$ &$4.66\cdot 10^6$ &$0.024$&$33.935$\\
 ($15$, $60$) & $27$ &$5$ &- &$2.97\cdot 10^{11}$ &$0.028$&-\\
 ($15$, $60$) & $28$ &$7$ &- &$3.95\cdot 10^{11}$ &$0.034$&-\\
 ($15$, $60$) & $34$ &$10$ &- &$2.86\cdot 10^{11}$ &$0.034$&-\\
 ($15$, $100$) & $46$ &$6$ &- &$9.46\cdot 10^{14}$ &$0.034$&-\\
 ($15$, $100$) & $48$ &$8$ &- &$9.35\cdot 10^{14}$ &$0.040$&-\\
 ($15$, $100$) & $51$ &$7$ &- &$9.35\cdot 10^{14}$ &$0.036$&-\\
 ($20$, $40$) & $19$ &$7$ &- &$1.18\cdot 10^{9}$ &$0.034$&$8\,663.920$\\
 ($20$, $40$) & $20$ &$10$ &- &$7.42\cdot 10^{8}$ &$0.044$&$5\,419.526$\\
 ($20$, $100$) & $47$ &$12$ &- &$5.76\cdot 10^{17}$ &$0.058$&-\\
 ($20$, $100$) & $48$ &$13$ &- &$4.43\cdot 10^{17}$ &$0.062$&-\\
 ($20$, $100$) & $52$ &$13$ &- &$4.15\cdot 10^{17}$ &$0.059$&-\\
 ($20$, $180$) & $83$ &$7$ &- &$8.91\cdot 10^{22}$ &$0.047$&-\\
 ($20$, $180$) & $84$ &$10$ &- &$8.89\cdot 10^{22}$ &$0.057$&-\\
 ($20$, $180$) & $103$ &$11$ &- &$8.94\cdot 10^{22}$ &$0.057$&-\\
 ($100$, $200$) & $93$ &$36$ &- &$4.94\cdot 10^{44}$ &$0.198$&-\\
 ($100$, $200$) & $101$ &$32$ &- &$5.06\cdot 10^{45}$ &$0.179$&-\\
 ($100$, $200$) & $102$ &$36$ &- &$1.06\cdot 10^{45}$ &$0.193$&-\\
 ($100$, $1\,000$) & $476$ &$39$ &- &$2.20\cdot 10^{125}$ &$0.296$&-\\
 ($100$, $1\,000$) & $488$ &$48$ &- &$5.74\cdot 10^{125}$ &$0.341$&-\\
 ($100$, $1\,000$) & $494$ &$50$ &- &$2.14\cdot 10^{125}$ &$0.345$&-\\
 ($100$, $2\,000$) & $981$ &$48$ &- &$3.00\cdot 10^{156}$ &$0.452$&-\\
 ($100$, $2\,000$) & $988$ &$43$ &- &$3.00\cdot 10^{156}$ &$0.427$&-\\
 ($100$, $2\,000$) & $1\,047$ &$52$ &- &$2.32\cdot 10^{156}$ &$0.457$&-\\
 ($100$, $4\,000$) & $1\,960$ &$49$ &- &$5.39\cdot 10^{186}$ &$0.700$&-\\
 ($100$, $4\,000$) & $1\,976$ &$46$ &- &$5.43\cdot 10^{186}$ &$0.681$&-\\
 ($100$, $4\,000$) & $1\,998$ &$55$ &- &$5.53\cdot 10^{186}$ &$0.723$&-\\
 ($1\,000$, $2\,000$) & $970$ &$355$ &- &- &$3.695$&-\\
 ($1\,000$, $2\,000$) & $976$ &$320$ &- &- &$3.445$&-\\
 ($1\,000$, $2\,000$) & $1\,008$ &$353$ &- &- &$3.695$&-\\
 ($1\,000$, $15\,000$) & $7\,419$ &$489$ &- &- &$8.101$&-\\
 ($1\,000$, $15\,000$) & $7\,441$ &$478$ &- &- &$7.999$&-\\
 ($1\,000$, $15\,000$) & $7\,541$ &$511$ &- &- &$8.325$&-\\
 ($1\,000$, $30\,000$) & $14\,894$ &$494$ &- &- &$12.805$&-\\
 ($1\,000$, $30\,000$) & $14\,947$ &$482$ &- &- &$12.416$&-\\
 ($1\,000$, $30\,000$) & $14\,988$ &$510$ &- &- &$12.708$&-\\
 ($1\,000$, $45\,000$) & $22\,430$ &$470$ &- &- &$17.375$&-\\
 ($1\,000$, $45\,000$) & $22\,548$ &$514$ &- &- &$17.996$&-\\
 ($1\,000$, $45\,000$) & $22\,632$ &$517$ &- &- &$18.011$&-\\
\end{tabular}}
\label{tab:numres2}
\end{table}

Table~\ref{tab:numres2} summarizes the times needed to compute all non-dominated outcome vectors with \esa\ for instances with up to $n=1000$ nodes and $m=45\,000$ edges. To randomly generate connected graphs, we use a code from \cite{schn:pkmax:2020} that first constructs a random spanning tree for the required number of nodes and afterwards the remaining edges are randomly added. 
For all instances, the cost coefficients of the first objective were uniformly distributed random integers between $1$ and $50\,000$ that were linearly transformed such that the smallest cost value is always equal to zero. For the second objective, the cost coefficients were uniformly distributed random integers from the set $\{0,1\}$. To reduce the effect of fluctuations due to varying processor loads, all times are averaged over ten runs on the same instance. 
Despite the exponentially growing cardinality $|\X|$ of the feasible set, which was computed using Kirchhoff's matrix tree theorem (see, e.g., \cite{russell:la:1994}) for instances up to $n=100$, the computational time of \esa\ always remains below one minute. 
It can be observed that both the number $|\Ynd|$ of non-dominated outcome vectors as well as the computational time needed by \esa\ grow mainly with $n$, and only marginally with $m$ and with the number $\vert E_1\vert$ of edges that have cost $1$, i.e., which are in the set $E_1=\{e\in\E:b(e)=1\}$ in the second objective $b$.

The numerical results shown in Table~\ref{tab:numres2} confirm the expected  efficiency  of \esa. Indeed, since  \esa\ computes the set of non-dominated outcome vectors $\Ynd$ (which has at most $n$ elements)  rather than the set of all efficient solutions $\Xeff$, the number of iterations of \esa\ is bounded by $n$. Moreover, each iteration requires a simple swap operation that can be implemented very efficiently.

Note that \esa\ generates only one pre-image, i.e., one feasible tree for each non-dominated outcome vector, while the number of efficient trees may be substantially larger. As an example, consider an instance where all edges have the same coefficients in both objectives. Then all spanning trees map to the same outcome vector in the objective space, i.e., $|\Ynd|=1$,  and are thus efficient, i.e., $|\Xeff|=|\X|$. 
In order to test whether this is a common situation also in randomly generated instances, we computed the complete set $\Xeff$ with the \ce\ approach for the smaller instances from  Table~\ref{tab:numres2}. It turns out that this is not the case for randomly generated instances on small graphs with a rather large range for the objective coefficients. 

Note also that since \esa\ exploits the fact that the non-dominated set $\Ynd$ solely consists of supported non-dominated outcome vectors when one of the objective functions has binary coefficients only (c.f.\ Lemma~\ref{t:bstpycon}), a numerical comparison with general solvers for bi- and multiobjective minimum spanning tree problems is not meaningful. In two-phase methods, for example, the search for unsupported non-dominated outcome vectors could be omitted, leading to an implementation that is somewhat similar to \esa.  On the other hand, algorithms that generalize classical methods for the single objective minimum spanning tree problem to the multiobjective case cannot be expected to be competitive with \esa\ since they generally enumerate far too many irrelevant trees. 

\subsubsection{Uniform matroids}
As a second test case we consider uniform matroids $\U_{k,n}$ on the ground set $\E=\{e_1,\dots,e_n\}$, from which exactly $k$ elements have to be selected in a basis. Rather than minimizing the cost of a basis we aim at maximizing its profit w.r.t.\ one general and one binary cost function to reflect the similarity of this problem to biobjective knapsack problems with bounded cardinality.
 
To determine the profit vectors of each element $e_i\in\E$, we generated $n$ uniformly distributed random values from the set $\{0,1,\dots,10n\}$ (for the first objective) and $n$ values from the set $\{0,1\}$ (for the second objective). After sorting the values for the first objective in non-decreasing order and the values of the binary objective in non-increasing order, the coefficients were combined into profit vectors for the elements $e_1,\dots,e_n$. 

For each instance on $n$ elements, Table~\ref{tab:numres3} shows the accumulated results over all values of $k\in\{1,\dots,\frac{n}{2}\}$. In order to analyse the relation between $|\Ynd|$ and $|\Xeff|$, we applied a simple implementation of a dynamic programming algorithm (\DP) for multiobjective knapsack problems as described, for example, in \cite{klam:dyna:00}. Different from the biobjective minimum spanning tree instances described above, we consistently observe that the number of efficient solutions exceeds the number of non-dominated outcome vectors, however, not by very much. As was to be expected, \esa\ easily solves larger instances within fractions of a second, while the computational time required by \DP\ grows significantly with the size of the instance. 
The efficency tests are run on a computer with an Intel(R) Core(TM) i7-7500U CPU @ 2.70GHz processor and 8 GB RAM. The algoithms are implemented in MATLAB, Version R2019b.

\begin{table}
    \centering
    \tbl{Computational results for randomly generated instances of uniform matroids $\U_{k,n}$. The two last columns show the accumulated average computation time over all $k=1,\dots,n/2$ in seconds, rounded over $10$ repetitions for each instance, for \esa\ (for the computation of $\Ynd$) and for \DP\ (for the computation of $\Xeff$), respectively. For instances with $80$ or more elements \DP\ needs more than $600$ seconds.}
{\begin{tabular}{r|r|r|r|r|r}
$n$ & $\vert E_1\vert$ & $|\Ynd|$ & $|\Xeff|$ & \esa\ [s] & \DP\ [s] \\
\hline
$20$ & $9$ & $64$ & $83$ & $0.02$ & $0.08$\\
$20$&$10$ &$65$ & $65$ & $0.02$ & $0.06$\\
$20$& $11$&$64$&$74$&$0.03$&$0.06$\\
$30$ & $12$ & $129$ & $129$ & $0.03$ & $0.44$\\
$30$ & $15$ & $135$ & $166$ & $0.03$ & $0.50$\\
$30$ & $16$ & $134$ & $134$ & $0.02$ &$0.30$\\
$50 $ & $21 $ & $340 $ & $364 $ & $0.06 $ & $35.33 $ \\
$ 50$ & $ 21$ & $340 $ & $340 $ & $0.07 $ & $27.29 $ \\
$50 $ & $29 $ & $340 $ & $413 $ & $0.06 $ & $14.43 $ \\
$60 $ & $ 27$ & $489 $ & $523 $ & $0.09 $ & $131.83 $ \\
$60 $ & $30 $ & $495 $ & $627 $ & $0.09 $ & $130.62 $ \\
$60 $ & $32 $ & $492 $ & $ 502$ & $0.09 $ & $80.36 $ \\
$70 $ & $ 34$ & $664 $ & $690 $ & $0.12 $ & $327.33 $ \\
$70 $ & $ 35$ & $665 $ & $716 $ & $0.12 $ & $336.23 $ \\
$ 70$ & $38 $ & $659 $ & $705 $ & $0.12 $ & $248.97$ \\
$80$ & $38$ & $857$ & $935$ & $0.21$ & $>600$\\
$80$ & $44$ & $850$ & $899$ & $0.27$ & $>600$ \\
$80$ & $50$ & $805$ & $860$ & $0.21$ & $>600$\\
$90$ & $45$ & $1\,080$ & $1\,181$ & $0.20$ & $>600$\\
$90$ & $46$ & $1\,079$ & $1\,213$ & $0.21$ & $>600$\\
$90$ & $47$ & $1\,077$ & $1\,159$ & $0.19$ & $>600$\\
$100$ & $46$ & $1\,315$ & $1\,361$ & $0.25$ & $>600$\\
$100$ & $50$ & $1\,325$ & $1\,483$ & $0.46$ & $>600$\\
$100$ & $51$ & $1\,324$ & $1\,441$ & $0.27$ & $>600$\\
\end{tabular}}
     \label{tab:numres3}
\end{table}

\subsection{Connectedness for more General Cost Functions}
The proof of the connectedness of the efficient set of Problem~\eqref{p:bbmp} (c.f.\  Theorem~\ref{cor:bbmpeffcon}) relies on two basic properties: On one hand, this is the matroid structure of the considered problem, and on the other hand it is the fact that one of the two objective functions has only binary cost coefficients. While the first property ensures the feasibility of elementary swap operations, the latter implies that two adjacent non-dominated outcome vectors always differ by exactly one unit in the binary objective function.

In general, i.e., when the objective coefficients can be chosen freely, biobjective optimization problems on uniform matroids may have non-connected efficient sets. Corresponding examples are provided in \cite{gors:conn:06} indicating that such non-connected instances (\emph{\nc-instances}) are very rare in randomly generated instances. The question remains whether non-connected instances already exist when the cost coefficients in the second objective are restricted to $\{0,1,2\}$ (rather than $\{0,1\}$), or, more generally, to $\{0,1,\dots,\beta\}$ with $\beta\geq 2$.

The frequency in which \nc-instances occurred for different values of $\beta$ in a large numerical study are reported in Table~\ref{tab:notbinary}. For each value of $\beta\in\{2,\dots,9,11,13,15,20,25,30\}$ we randomly generated $300\,000$ instances of uniform matroids $\U_{k,n}$ with $n=20$ elements. The profit vectors were chosen as described in Section~\ref{subsec:numESA} above, where the coefficients for the second objective were now drawn from the set $\{0,1,\dots,\beta\}$. 
All instances were solved for all $k\in\{1,\dots,\frac{n}{2}\}$ using a \DP\ approach for multiobjective knapsack problems, see \cite{klam:dyna:00}.

\begin{table}
     \centering
     \tbl{Number of observed  \nc-instances in $300\,000$ randomly generated instances of $\U_{k,20}$, cumulated for all $k\in\{1,\dots,10\}$, for different values of $\beta$.}
{\begin{tabular}{
r||c|c|c|c|c|c|c|c|c|c|c|c|c|c}
$\beta$ & $2$ & $3$ &$4$ & $5$ & $6$ & $7$ & $8$ &$9$ &$11$ &$13$ &$15$ &$20$ &$25$ &$30$  \\
\hline
 \nc-instances & $0$ & $0$ &$0$&$0$ &$0$ &$0$&$2$ &$4$ & $3$ &$2$ &$8$ &$10$ &$12$ &$13$
 \end{tabular}}
\label{tab:notbinary}
 \end{table}
 
Table~\ref{tab:notbinary} indicates that it seems to become more likely to find \nc-instances the larger the range for the coefficients in the second objective function is, i.e., the larger the value of $\beta$ is. 
Nevertheless, we suspect that \nc-instances also exist for smaller values of $\beta$ but that such instances are extremely rare. While \nc-instances may be more likely for larger values of $n$, analysing large data sets becomes more and more challenging since this requires the exact computation of the complete efficient set for each instance (without the possibility of using \esa). We note that preliminary tests with $n=30$ and $n=50$ did not provide further insight on this topic.

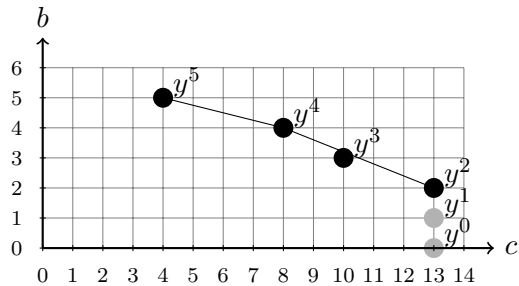
\begin{figure}[tbh]
\begin{center}
  \begin{tikzpicture}[scale=0.4]
\draw[very thin,color=gray] (0,0) grid (14,6);
\fill[fill=black!30] (13,0) circle (2ex);
\node (A) at (13.8,0.5) [circle] {$y^0$};
\fill[fill=black!30] (13,1) circle (2ex);
\node (B) at (13.8,1.5) [circle] {$y^1$};
\fill (13,2) circle (2ex);
\node (C) at (13.8,2.5) [circle] {$y^2$};
\fill (10,3) circle (2ex);
\node (D) at (10.8,3.5) [circle] {$y^3$};
\fill (8,4) circle (2ex);
\node (E) at (8.8,4.5) [circle] {$y^4$};
\fill (4,5) circle (2ex);
\node (F) at (4.8,5.5) [circle] {$y^5$};
\draw[-] (13,2) to (8,4);
\draw[-] (8,4) to (4,5);

\draw[->,thick] (0,0) -- (15,0) node[right] {$c$};
\draw[->,thick] (0,0) -- (0,7) node[above] {$b$};
\foreach \x in {0,1,2,3,4,5,6,7,8,9,10,11,12,13,14}\draw (\x,-0.1) -- (\x,0.1) node[below=5pt] {$\scriptstyle\x$};
\foreach \y in {0,1,2,3,4,5,6}\draw (-0.1,\y) -- (0.1,\y) node[left=5pt]  {$\scriptstyle\y$};
\end{tikzpicture} 
\end{center}
\caption{Non-dominated set of the uniform matroid introduced in Table~\ref{tab:nonsupported1}. Non-dominated outcome vectors are shown in black and weakly non-dominated points shown in grey. The point $y^{3}$ corresponds to a non-dominated and non-supported outcome vector.} \label{f:nonsupported}
\end{figure}

A necessary condition for the existence of nc-instances is the existence of non-dominated non-supported outcome vectors. But in contrast to nc-instances it is quite easy to generate knapsack problems with such outcome vectors. An example  is given in Table~\ref{tab:nonsupported1}. In the first column of the left table the number of an item, in the second column the first weight and in the third column the second weight with $\beta=2$ is given. In the right table, the efficient bases for $k=3$ are given with their outcome vectors. Recall that we interpret the uniform matroid as a special case of a knapsack problem with bounded cardinality and thus consider both objective functions as maximization objectives. As can be seen in  Figure~\ref{f:nonsupported}, the basis $\{e_1,e_4,e_5\}$ is non-dominated and non-supported. Nevertheless, the efficient set of this instance is connected.

\begin{table}
     \centering
     \tbl{Left: Elements $e_i$ $i=1,\dots,k$ of a knapsack $\U_{3,6}$ with their weights with $\beta=2$. Right: Efficient bases for the knapsack.}
{\begin{tabular}{
c|c|c c c|c|c}
$e$ & $c(e)$ & $b(e)$& \hspace*{1.5cm} &$B$ & $c(B)$ & $b(B)$   \\
\cline{1-3}\cline{5-7}
$e_1$ & $6$ & $0$ & &$\{e_1,e_2,e_3\}$ & $13$ & $0$ \\
$e_2$ & $5$ & $0$ & &$\{e_1,e_2,e_4\}$ & $13$ & $1$\\
$e_3$ & $2$ & $0$ & &$\{e_1,e_2,e_5\}$ & $13$ & $2$\\
$e_4$ & $2$ & $1$ & &$\{e_1,e_4,e_5\}$ & $10$ & $3$\\
$e_5$ & $2$ & $2$ & &$\{e_1,e_5,e_6\}$ & $8$ & $4$\\
$e_6$ & $0$ & $2$ & &$\{e_4,e_5,e_6\}$ & $4$ & $5$\\
 \end{tabular}}
\label{tab:nonsupported1}
 \end{table}
 
\section{Conclusions}\label{sec:bbmpcon}

In this paper we investigate biobjective matroid problems involving one binary cost objective. We present an efficient swap algorithm (\esa) that solves this special kind of biobjective matroid problem efficiently, although the decision problem of the general version of this problem is known to be NP-complete (cf.\ \cite{ehrg:onma:96}).  The idea of \esa\ is based on a method of \cite{gabo:effi:84} for a constrained version of single-objective matroid optimization problems. The complexity of \esa\ depends on the matroid type. For a graphic matroid on a graph $G=(V,E)$, for example, it is given by $\bigo(m+n\cdot\log(n))$, where $|V|=n$ and $|E|=m$. Numerical experiments confirm the efficiency of this approach.

The efficient swap algorithm can be interpreted as a neighborhood search approach with an efficient strategy for the identification of relevant swaps. The correctness of this approach is based on the proof of the connectedness of the efficient set in this special case, which is in turn based on the insight that the non-dominated set consists only of supported non-dominated outcome vectors. This is surprising since  it was shown in \cite{gorski:diss:2010} that the efficient set is in general  non-connected for biobjective matroid problems. To the best of our knowledge this is the first class of problems where connectedness of $\Xeff$ can be established even though the non-dominated set is not contained in a hyperplane.

\bibliographystyle{tfnlm}
\bibliography{matroid}

\end{document}